\documentclass[a4paper,fleqn,usenatbib]{mnras}

\usepackage{newtxtext,newtxmath}
\usepackage{color}
\usepackage[T1]{fontenc}
\usepackage{ae,aecompl}

\usepackage{graphicx,soul}	% Including figure files
\usepackage{amsmath}	% Advanced maths commands
\usepackage{amssymb}	% Extra maths symbols

\newcommand{\ud}{\,\mathrm{d}}

\newcommand{\lsat}{L_{\mathrm{sat}}}
\newcommand{\reff}{R_{\mathrm{eff}}}
\newcommand{\msun}{M_\odot}
\newcommand{\msol}{M_\odot}

\title[SDSS central galaxy properties and their host halos]{Connecting SDSS central galaxies to their host halos using total satellite luminosity}

\author[Alpaslan \& Tinker]{
Mehmet Alpaslan$^{1}$\thanks{E-mail: mehmet.alpaslan@nyu.edu (MA)} and 
Jeremy L. Tinker$^{1}$
\\
$^{1}$Center for Cosmology and Particle Physics, Department of Physics, New York University, New York, NY 10012, USA
}

\date{Accepted XXX. Received YYY; in original form ZZZ}

\pubyear{2019}

\hypersetup{draft}
\begin{document}
\label{firstpage}
\pagerange{\pageref{firstpage}--\pageref{lastpage}}
\maketitle

% Abstract of the paper
\begin{abstract}
The total luminosity of satellite galaxies around a central galaxy, $\lsat$, is a powerful metric for probing dark matter halos. In this paper we utilize data from the Sloan Digital Sky Survey and DESI Legacy Imaging Surveys to explore the relationship between $\lsat$ and various observable galaxy properties for a sample of 117,966 central galaxies out to $z = 0.15$. At fixed stellar mass, we find that every galaxy property we explore shows a correlation with $\lsat$. This implies that dark matter halos play a role---possibly a very significant role---in determining these secondary galaxy properties. We quantify these correlations by computing the mutual information between $\lsat$ and secondary properties, explore how this mutual information varies as a function of stellar mass and when separating the sample into star-forming and quiescent central galaxies. We find that absolute r-band magnitude correlates more strongly with $\lsat$ than stellar mass across all galaxy populations; and that effective radius, velocity dispersion, and S\'ersic index do so as well for star-forming and quiescent galaxies. The $\lsat$ observable is sensitive to both the mass of the host halo as well as the halo formation history, with younger halos having higher $\lsat$. $\lsat$ by itself cannot distinguish between these two effects, but measurements of galaxy large-scale environment can break this degeneracy. For star-forming central galaxies, we find that $\reff$, $\sigma_v$, and S\'ersic index all correlate with large-scale density, implying that these halo age plays a role in determining these properties. For quiescent galaxies, we find that all secondary properties are independent of environment, implying that correlations with $\lsat$ are driven only by halo mass. These results are a significant step forward in quantifying the full extent of the galaxy-halo connection, and present a new test of galaxy formation models. 

\end{abstract}

% Select between one and six entries from the list of approved keywords.
% Don't make up new ones.
\begin{keywords}
keyword1 -- keyword2 -- keyword3
\end{keywords}

%%%%%%%%%%%%%%%%%%%%%%%%%%%%%%%%%%%%%%%%%%%%%%%%%%

%%%%%%%%%%%%%%%%% BODY OF PAPER %%%%%%%%%%%%%%%%%%

\section{Introduction}
\label{sec:intro}

Galaxies are profoundly influenced by the dark matter halo within which they are embedded, particularly within the context of processes that regulate star formation. Vital star forming gas can be stripped from galaxies as they infall into groups \citep{Kimm2009,Weinmann2010}, leading to observable differences between central and satellite galaxies \citep{Peng2012,Darvish2016}. At much larger scales, galaxies can experience a similar suppression of their star formation rates as they fall onto large-scale filaments \citep{Alpaslan2016,Aragon-Calvo2016a,Kraljic2018}.  Concurrently, processes internal to a galaxy can have equally impactful effects on its observable properties, including its star formation rates. It is therefore crucial to be able to accurately determine the environment a galaxy inhabits in order to be able to differentiate between the instrinsic and extrinsic effects that shape its evolution. Yet despite the pivotal role that dark matter haloes play in galaxy evolution, quantifying properties of the halo itself remains a  challenge observationally.  One common method for identifying which galaxies are central and which are satellites, as well as the masses of the halos they reside in, is galaxy group finders applied to data from spectroscopic surveys (e.g. \citep{Yang2005,Tinker2011,Robotham2011,Alpaslan2012}. Accurate determinations of galaxy groups are vital for estimating the form of the galaxy-halo connection, which in turn provide important constraints both for models of galaxy formation and evolution, as well as as cosmology (see e.g. \citealp{Wechsler2018}).

In one sense, the relationship between galaxies and halos is abundantly clear: more massive galaxies live in more massive halos, a consequence of the larger reservoir of baryons from which stars can form (see e.g. \citealp{Kravtsov2013} and \citet{Huang2017}). But even at fixed stellar mass, galaxies show a wide diversity of properties: different sizes, morphologies, kinematics, and stellar populations. All these variations reflect the myriad processes that come into play in the physics of galaxy formation. But one general question---one that has not been answered---is the role of the dark sector in creating this galactic diversity. For galaxies at the same stellar mass, are the differences in---for example---stellar velocity dispersion driven purely by baryonic physics or is it correlated with the properties of the host halos these different galaxies live in? In this paper, we present a test that provides a significant step forward in answering this question.

While galaxy group catalogues and other, more direct techniques for measuring dark matter halo masses such as gravitational lensing (e.g. \citealp{VanUitert2016,Mandelbaum2016}) have been successful in studies of galaxy evolution, they are limited by signal-to-noise constraints, limiting measurements to $M_h \gtrsim 10^{12}$ $\msol$, and preveting fine-binning of samples by galaxy properties. Recently, in Tinker et al. (in prep; hereafter T19) we combined spectroscopic and deep photometric data to show that the total luminosity of satellite galaxies around a central galaxy can probe the properties of dark matter halos. Hereafter referred to as $\lsat$, the total satellite luminosity for a given central galaxy is measured by integrating the background-subtracted luminosity from satellite galaxies that surround it. By combining N-body simulations with abundance matching models, T19 are able to show that $\lsat$ scales linearly with the host halo mass of the central galaxy, and has a strong relationship with the stellar mass of the central galaxy. This relationship holds even when measuring $\lsat$ within projected apertures of fixed radius (e.g. 50 or 100 h$^{-1}$ kpc). But more than just being a probe of halo mass, $\lsat$ is also sensitive to the formation history of the halo, with younger halos having more substructure, and thus more faint satellite galaxies. T19 demonstrated that $\lsat$, combined with measurements of large-scale galaxy environment, can break this degeneracy and detect correlations with {\it both} halo mass and halo age. Based on these results, $\lsat$ is an ideal proxy for measuring dark matter halo properties, yielding high signal-to-noise results even for galaxies residing in halo as low as $10^{11}$ $\msol$.

The motivation behind using large-scale environment to break this degeneracy springs from the fact that halos exhibit assembly bias: at fixed halo mass, certain secondary properties of the dark matter halo show a correlation with large-scale environment. Measuring the clustering of halos will therefore pick up on these biases. The amount of substructure in a halo is one such secondary parameter; and we know that halos that form at earlier time have fewer numbers of subhalos within them\footnote{Subhalos within early-forming halos have a much longer amount of time to merge with their parent halo due to dynamical friction or mergers; whereas late-forming halos have more recent mergers with subhalos, meaning that these substructures remain distinguishable and discrete.}. Halo formation history itself correlates with density, with late-forming halos preferentially residing in low density regions, and these late-forming halos typically have lower concentrations than those that form early. 

Recent observational studies of assembly bias have added to our growing understanding of the myriad of physical processes that can influence halo properties; with no single halo property being solely responsible for the observed bias of halos \citep{Mao2018}. In fact, no single halo property can even account for the bias in the spatial clusteing of dark matter halos \citep{Salcedo2017}. On the other hand \citet{Han2019} argue that this complex multivariate relationship between internal halo properties and assembly bias is, itself, dependent on the mass of the halo; environmental density on its own can account for approximately 30\% of halo bias effects. \citet{Ramakrishnan2019} further show that a dark matter halo's cosmic web environment can play a significant role in its bias. The work presented in this manuscript is therefore well placed to contribute to the growing understanding of the role that environment plays in shaping halo properties; particularly as the satellite population of a central galaxy is explicitly sensitive to the environment the halo is found in.

In this manuscript we explore the relationship between $\lsat$ for central galaxies stacked on eight different observable parameters at fixed stellar mass. Given the strong link established between $M_h$, $M_*$ and $\lsat$ in T19, it is important to examine which galaxy properties besides stellar mass correlate with $\lsat$, and how these correlations change at fixed stellar mass. We additionally examine the relationship between these parameters and the large-scale density of galaxies to break the aforementioned degeneracy between halo mass and halo age in influencing the relationship these parameters have with $\lsat$. These relationships, once understood, can then be used to better constrain galaxy formation models, as well as halo mass estimates of galaxy groups made using observational data sets; a topic that we will address in a future paper. This paper is organized as follows: in Section \ref{sec:lsat} we introduce our data and describe the total satellite luminosity, or $\lsat$ parameter in greater detail, including a discussion on how we compute this parameter for our sample of galaxies. We additionally investigate the global properties of $\lsat$ and systematics in our measurement. Section \ref{sec:properties} investigates the relationships between different galaxy observables and $\lsat$ and examines how these correlations perform better than correlations between $\lsat$ and stellar mass. In Section \ref{sec:discussion} we discuss and interpret our results, and conclude with a summary in Section \ref{sec:conclusion}. Throughout this work we adopt the following cosmological parameters: $H_0 = 70$ km s$^{-1}$, $\Omega_M = 0.25$, and $\Omega_L = 0.75$. Unless otherwise noted, all stellar masses and luminosities are given in units of solar masses and luminosities.

\section{Sample selection and luminosity calculation}
\label{sec:lsat}

\subsection{Finding central galaxies}

Central galaxy identification is typically done using groupfinding algorithms (e.g. \citealp{Yang2005,Tinker2011,Robotham2011}), or isolation criteria (e.g. \citealp{Kauffmann2010}). In this work we use the central finding algorithm of T19. This central finder is introduced and discussed in detail in T19, as accurate determination of central galaxies is the biggest source of uncertainty in $\lsat$. Here we provide an abridged description of this algorithm, and poinst the reader to T19 for a more detailed discussion. For a given galaxy, the probability that it is the central galaxy of the halo it belongs to is given by:

\begin{equation}
	P_{\mathrm{cen}} = \frac{1}{1 + (P_{R_p} P_{\Delta_z} / B)}
	\label{eqn:psat}
\end{equation}

\noindent where $P_{R_p}$ is the probability that a galaxy is a satellite by given its projected distance from the center of the halo assuming a standard NFW profile \citep{Navarro1997}; and $P_{\Delta_z}$ is the same probability but given the line-of-sight separation from the center of the halo assuming a Gaussian velocity distribution. In the case of $P_{R_p}$ the probability is computed assuming the halo is populated with galaxies whose number density profile follows that of a standard NFW profile. For $P_{\Delta_z}$ we assume that the line-of-sight number density profile of galaxies about the center of the halo follows a standard Gaussian distribution. $B$ is a scaling constant determined from calibrations on mock galaxy samples and is set to be 10. For additional details of this procedure refer to \citet{Yang2005,Tinker2011}. The mass of the halo that a galaxy is within is determined using pre-tabulated stellar to halo mass relationships between $0 < z < 8$ from \citet{Behroozi2013}, removing the need to use a volume limited sample of galaxies (thus ensuring a larger sample size, reducing the noise on $\lsat$). On the other hand, this central finding algorithm differs from the groupfinder used by \citet{Tinker2011} in that it does not iterate towards a stable group halo mass, which leads to some impurities in the resulting population of central galaxies in the form of misclassified satellite galaxies. Uncertainties associated with these misclassifications, however, are less significant than the noise in $\lsat$ that would come about from using a volume-limited catalogue with a significantly lower number of galaxies. For the purposes of this work we only consider galaxies with a satellite probability $P_{\mathrm{sat}} =  P_{\mathrm{cen}} - 1 \leq 0.1$ to be central galaxies. This extremely conservative cut to the satellite probability of a galaxy ensures that we are not at risk of misclassified satellite galaxies contaminating our results.

The central galaxies in our study are selected with spectroscopic data from the Main Galaxy Sample of the Sloan Digital Sky Survey \citep{Strauss2002}; specifically utilizing the NYU-VAGC catalogues from \citet{Blanton2005}. The redshift distribution of central galaxies peaks at $z = 0.0735$ and extends past $z = 0.15$; however, we only consider galaxies with $z \leq 0.15$ in this work as some of the photometric galaxy properties we examine are taken from the NASA Sloan-Atlas catalog \citep{Blanton2011}\footnote{http://nsatlas.org/} (hereafter NSA catalogue), which has an upper redshift limit of $z = 0.15$.

\subsection{Computing $\lsat$}

\begin{figure}
	\centering
	\includegraphics[width=0.5\textwidth]{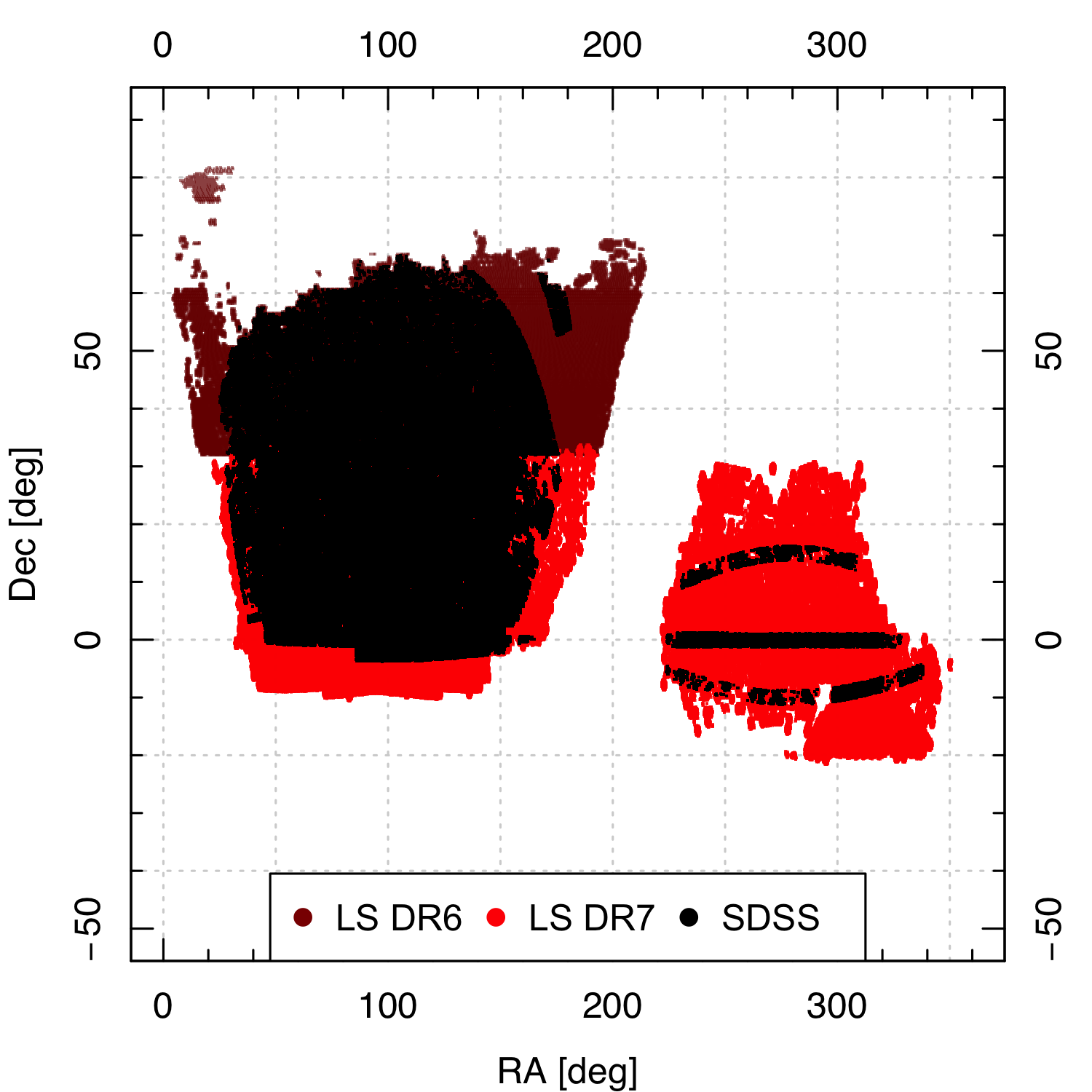}
	\caption{Right ascension (rotated by 85 degrees) and declinations of SDSS central galaxies used in this study, plotted as black points. The photometric data from the DESI Legacy Imaging Surveys that we use to compute $\lsat$ completely overlaps this region, and is shown in dark red (DR6) and red (DR7).}
	\label{fig:skymap}
\end{figure}

Total satellite luminosities around each central galaxy are calculated using imaging data from a combined sample of imaging data from three surveys referred to collectively as the DESI Legacy Imaging Surveys: the DECam Legacy Survey (DECaLS; \citealp{Flaugher2015}), Beijing-Arizona Sky Survey (BASS; \citealp{Zou2017}), and Mayall z-band Legacy Survey (MzLS; \citealp{Zhou2018}). For this work we use photometry from the 6th and 7th data releases of the DESI Legacy Imaging Surveys. While these surveys are ongoing, their current extant footprint nearly overlaps the region of the sky from which our spectroscopic central galaxy sample is drawn (see Figure \ref{fig:skymap}). We filter these photometric data to exclude any sources that are considered point sources; lack imaging in any one of the $g$, $r$, or $z$ bands; have negative flux variance; and are more than 60\% masked. This selection ensures a clean and complete photometric sample for $\lsat$ calculations. We only compute satellite luminonsities for spectroscopic SDSS central galaxies that are more than $R = 3R_{\mathrm{vir}}$ from the DESI Legacy Imaging Surveys edge. We exclude all SDSS central galaxies that are within 50 h$^{-1}$ kpc of another SDSS central galaxy (as these would count as interlopers when estimating the background luminosity $L_{\mathrm{BG}}$) -- this only reduces our sample size by approximately 6\%.

A full description of how $\lsat$ is computed in simulated data as well as in this sample of SDSS galaxies is given in T19, including analysis of how $\lsat$ varies as a function of different aperture sizes (i.e. beyond 50 h$^{-1}$ kpc). Here we provide an abridged description of how $\lsat$ is measured for SDSS galaxies using LS data. For a given central galaxy, $\lsat$ is defined to be:

\begin{equation}
\lsat = \sum_{M_{\mathrm{r}}^{\mathrm{low}}}^{M_{\mathrm{r}}^{\mathrm{hi}}} 10^{0.4(M_{\mathrm{r}} - M_{\mathrm{r},\,\odot})} \Phi(M_{\mathrm{r}})\Delta M_{\mathrm{r}}
\label{eqn:lsat}
\end{equation} 

\noindent where $\Phi(M_{\mathrm{r}})$ is the conditional luminosity function, which is defined as the luminosity of satellite galaxies around a halo of a given mass. More explicitly, this is defined as:

\begin{equation}
	\Phi_{\mathrm{sat}} (M_{\mathrm{r}} | M_{\mathrm{h}}) \ud M_{\mathrm{r}} = \frac{N_{\mathrm{tot}}(M_{\mathrm{r}}) - N_{\mathrm{BG}}(M_{\mathrm{r}})}{N_{\mathrm{h}}(M_{\mathrm{r}})}
	\label{eqn:CLF}
\end{equation}

\noindent where $N_{\mathrm{tot}}$ is the total number of galaxies with a given limited magnitude $M_{\mathrm{r}}$ in a circular aperture around the central galaxy, and $N_{\mathrm{BG}}$ is the number of background galaxies at that magnitude; similar to the methodology employed by \citet{Hansen2007,Tal2012} (see T19 for additional details on computing $N_{\mathrm{BG}}$). $N_{\mathrm{h}}$ is the number of halos where the limiting magnitude is M$_{\mathrm{r}}$ or higher. Returning to Eq. \ref{eqn:lsat}, we compute $\lsat$ for each galaxy by integrating from the limiting magnitude to $M_r^{\mathrm{hi}}$, with $M_{\mathrm{r},\,\odot} = 4.65$. ${M_{\mathrm{r}}^{\mathrm{hi}}}$ is chosen based on the stellar mass of the galaxy to ensure that the satellite galaxies are not brighter than the central galaxy itself; ${M_{\mathrm{r}}^{\mathrm{hi}}} = -21 - 2\times(\log M_* - 10)$. $M_{\mathrm{r}}^{\mathrm{low}}$ is chosen to be $-14$. Naturally, there is some contribution to the $\lsat$ value computed in Eq. \ref{eqn:lsat} from background sources. We estimate this contaminant background luminosity, $L_{\mathrm{BG}}$ by calculating the average total luminosity contained in randomly placed 50 kpc apertures that are outside of $R_{\mathrm{vir}}/2$ of any central SDSS galaxy. We remove this contaminant background value from all reported $\lsat$ measurements. See Figure \ref{fig:skymap} for an Aitoff projection sky map of the SDSS central galaxies used in this study, as well as the coverage areas of the DESI Legacy Imaging Surveys.

\subsection{Computing galaxy densities}

\begin{figure}
	\includegraphics[width=0.5\textwidth]{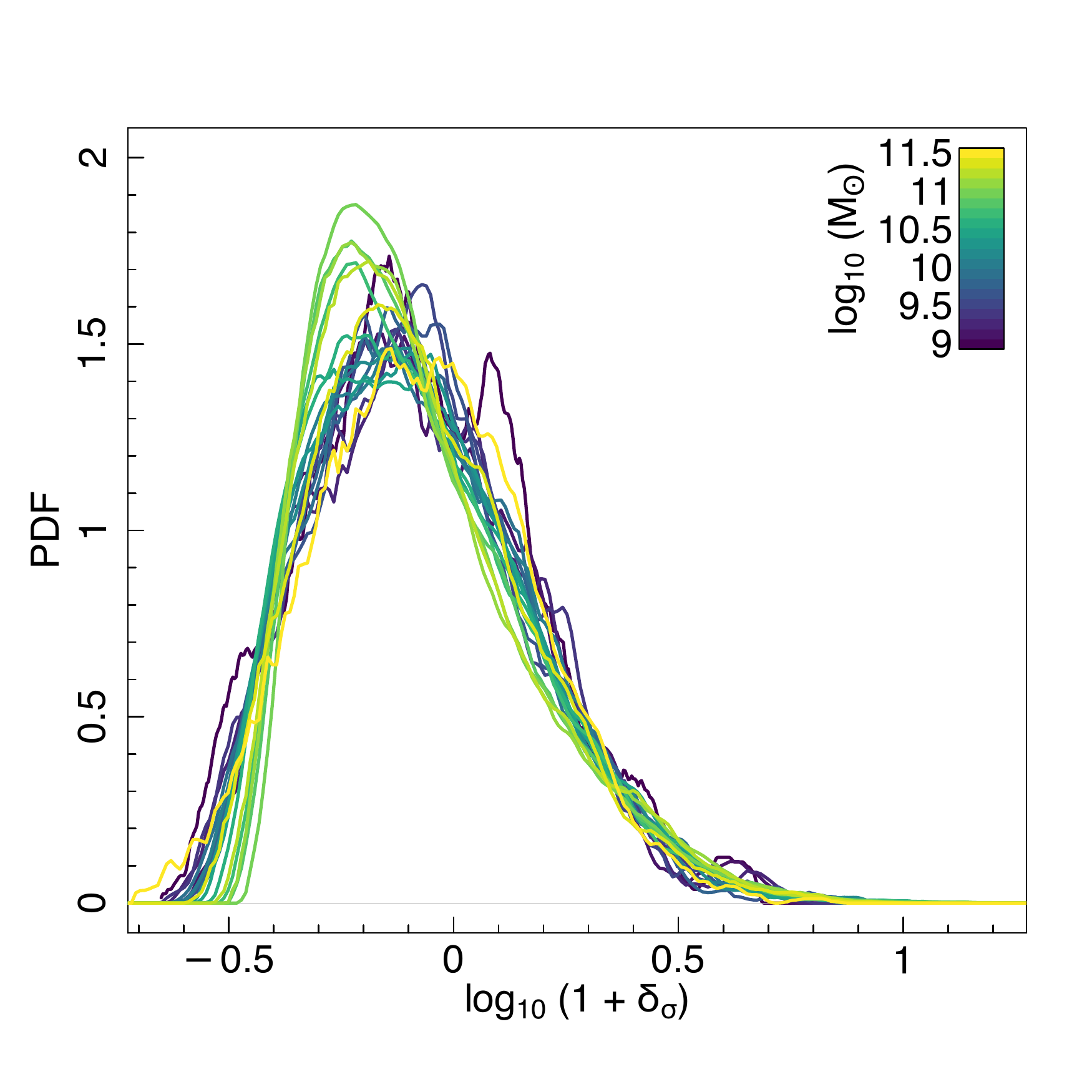}
	\caption{PDFs of $log_{10} (1 + \delta_{\sigma_z})$ computed in bins of fixed stellar mass, shown in different colours according to the colour bar on the top left. The distribution of $\delta_{\sigma_z}$ does not change significantly as a function of stellar mass.}
	\label{fig:densPDFs}
\end{figure}

As discussed in the introduction, halo formation history correlates strongly with density; and $\lsat$ is sensitive to halo formation history. To further study this relationship, we wish to use a metric of density that has a maximal signal-to-noise ratio at all redshifts, but is independent of redshift. Were we to compute galaxy densities using a density-defining population (DDP), we would be biased towards brighter objects in order to ensure our DDP was volume limited out to $z = 0.15$. Instead, we devise a density metric $\delta_{\sigma}$ such that for each central galaxy

\begin{equation}
	\delta_{\sigma} = \frac{\rho_z - \langle \rho_z \rangle}{\sigma_z}
	\label{eqn:dens}
\end{equation}

\noindent where $\rho_z$ is the number density of SDSS galaxies with respect to a background galaxy population within 10 Mpc centered on each galaxy at fixed redshift; $\langle \rho_z \rangle$ is the mean value of $\rho_z$ for a given bin of galaxies (when binned by redshift or stellar mass, for example); and $\sigma_z$ is the root-mean-square of $\rho_z$ for that same binned population of galaxies. We utilize this density metric rather than a more straightforward one because we are not using a volume limited background galaxy sample for our number count (doing so would reduce the number of background galaxies we can count from, introducing larger errors in our density measurements). Figure \ref{fig:densPDFs} shows that the distribution of $\delta_{\sigma_z}$ does not vary greatly as a function of stellar mass.

For additional details pertaining to how $\delta_\sigma$ is computed (including how we account for edge effects) as well as diagnostic plots see Appendix \ref{sec:densApp}.
 
\section{Total satellite luminosity as a function of galaxy properties}
\label{sec:properties}

In this section we will examine the relationship between $\lsat$ computed within 50 h$^{-1}$ kpc around each central galaxy and eight of its observable properties discussed above. Our final sample contains 117,966 central galaxies out to $z = 0.15$. We apply a k-correction to the $\lsat$ value of each galaxy by converting this total luminosity to a magnitude value, to which we then apply a k-correction at the redshift of the central galaxy using \textsc{Kcorrect v4.2} \citep{Blanton2006}. This k-corrected magnitude is then converted back to a luminosity value. Finally, each galaxy's $\lsat$ estimate is corrected for incompleteness by upweighting it using $1/v_{\mathrm{max}}$ estimates taken from the NSA catalogue, though this has very little impact on the values of $\lsat$.

In this work, we are interested in examining how $\lsat$ varies as a function of eight galaxy properties: stellar mass ($M_*$; $\log_{10}$ M$_{\odot}$); effective radius ($\reff$; h$^{-1}$ kpc); velocity dispersion ($\sigma_v$; km s$^{-1}$); concentration (c$_{90/50}$); $r$-band absolute magnitude ($M_r$; mags); 4000 \r{A}ngstrom break (D$_n 4000$); specific star formation rate (sSFR; yr$^{-1}$); surface mass density ($\Sigma_*$; $\log_{10}$ M$_{\odot}$ kpc$^{-2}$ h$^2$); and S\'ersic index ($n_s$). All galaxy properties are taken from the NYU-VAGC catalogue, the NSA catalogue and the MPA-JHU catalogue \citep{Brinchmann2004}. We utilize stellar masses derived from applying principal component analysis to optical rest-frame SDSS spectra, referred to as `PCA stellar masses' \citep{Chen2011}.

\subsection{Central galaxy properties and $\lsat$ as a function of large-scale density}

\begin{figure}
	\centering
	\includegraphics[width=0.5\textwidth]{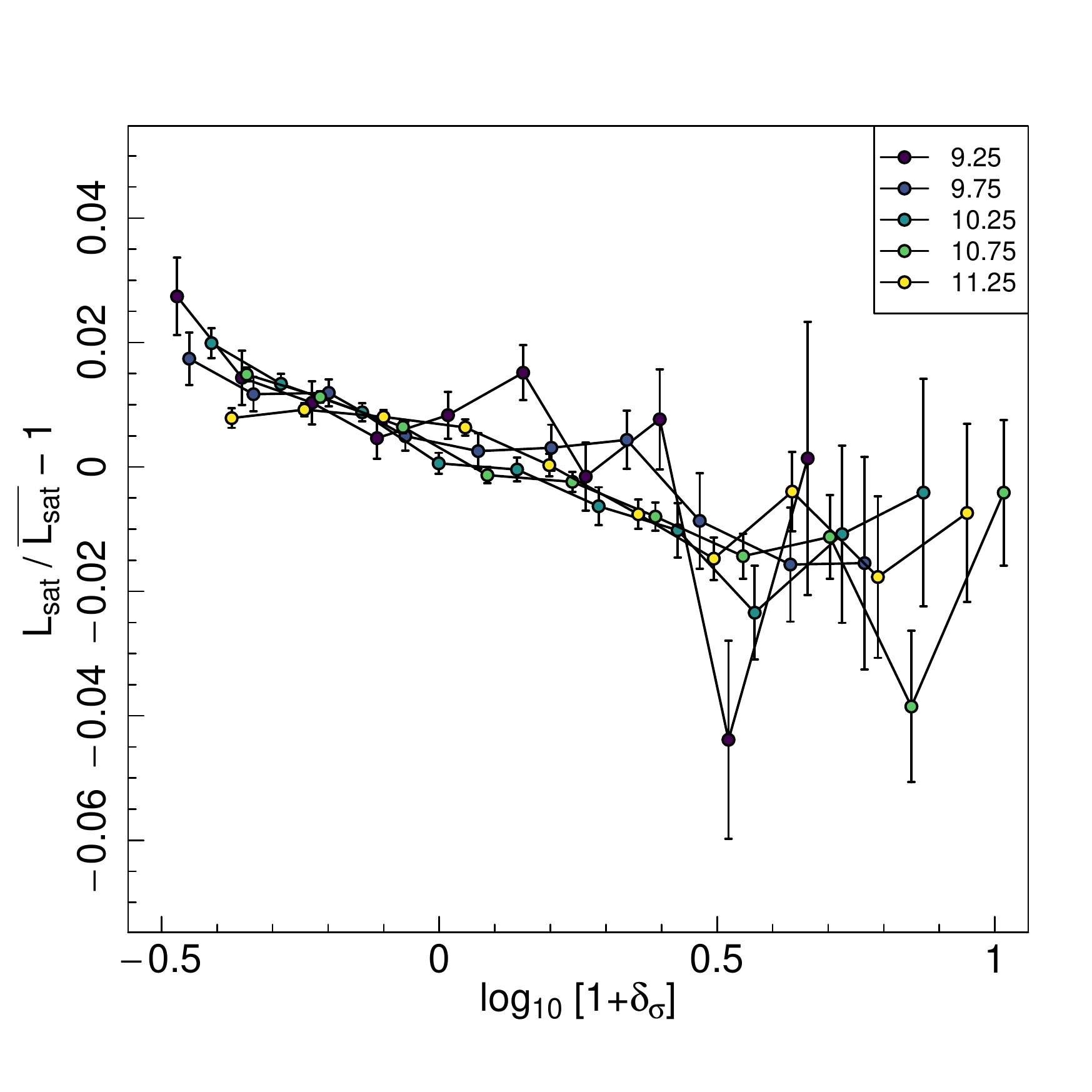}
	\caption{$\lsat / \bar{\lsat} - 1$ shown as a function of $\log_{10} (1 + \delta_\sigma)$ for galaxies binned by stellar mass, as shown by the key in the upper right. Points plotted are median values for each bin, and the error bars show the standard error about the median. Across all stellar mass values there is very little variation in how $\lsat$ behaves as a function of density. We do not see significant differences in the relationship between $\lsat$ and density when we split galaxies by colour using their D$_n4000$ values.}
	\label{fig:lsatdens}
\end{figure}

In Figure \ref{fig:lsatdens} we examine how $\lsat$ behaves as a function of density across the stellar mass range of our sample of galaxies. On the $y-$axis we plot a value we refer to as $\lsat / \bar{\lsat} - 1$. $\bar{\lsat}$ is the $\lsat$ value of a galaxy inferred from its stellar mass by fitting a power law to the $\lsat - M_*$ relationship for galaxies in each mass bin. $\lsat / \bar{\lsat}$ therefore effectively measures the offset between a galaxy's $\lsat$ value and the mean $\lsat$ value of galaxies at that stellar mass. Performing this computation ensures our results are unbiased with respect to inherent trends in $\lsat$ with stellar mass, even at fixed stellar mass (we explore these trends later in the paper). We plot the mean value of this ratio for each stellar mass bin as a function of $1 + \log_{10}(1+\delta_\sigma)$. Figure \ref{fig:lsatdens} shows that there is no systematic change in the relationship between $\lsat$ and density at fixed redshift and stellar mass. While it is true that galaxies with higher values of $\lsat$ tend to be found at lower densities (modulo some significant scatter at high densities), this relationship holds true across all mass ranges that we explore. Our results are therefore not affected systematically by changes in the $\lsat$-density relationship as a function of redshift and mass. We note that though there is a slope in the relationship between $\lsat$ and density, this is only of order 2\%. This figure also shows that our central galaxies are well identified: T19 show that satellite galaxies have significantly higher values of $\lsat$ compared to central galaxies. If we had been significantly misclassifying satellite galaxies as central galaxies, this would be more likely to happen at high values of density. The fact that we do not see an upward trend in $\lsat$ as a function of density therefore means that we are correctly identifying central galaxies most of the time. The lack of this upward trend also suggests that there are no significant numbers of interloper SDSS central galaxies within the line of sight of our sample central galaxies (as this would also drive $\lsat$ to higher values at high densities). Finally, we note that the high values of $\log_{10} (1 + \delta_\sigma)$ seen in Figure \ref{fig:lsatdens} are not entirely unexpected, as density follows a log-normal distribution with very large tails.

As stated in the introduction and addressed in T19, any correlation of $\lsat$ with a secondary galaxy property may be driven by two separate correlations: with halo mass, or halo formation history (late-forming halos will have higher $\lsat$ as they have not had time to accrete infalling galaxies into the central). These two correlations, however, make distinct predictions on how each property should correlate with $\delta_\sigma$, especially at M$_h \leq 10^{12.5}$ M$_\odot$, where halo clustering becomes independent of halo mass. T19 further show explicitly that it is possible to distinguish whether or not the mass or formation history of a halo is driving the correlation between a secondary galaxy parameter and $\lsat$ by looking at how that secondary galaxy parameter varies as a function of large-scale density. The relationship between $\lsat$ and a galaxy property that does not vary as a function of large-scale density is driven entirely by $M_h$.

\begin{figure*}
	\centering
	\includegraphics[width=1.0\textwidth]{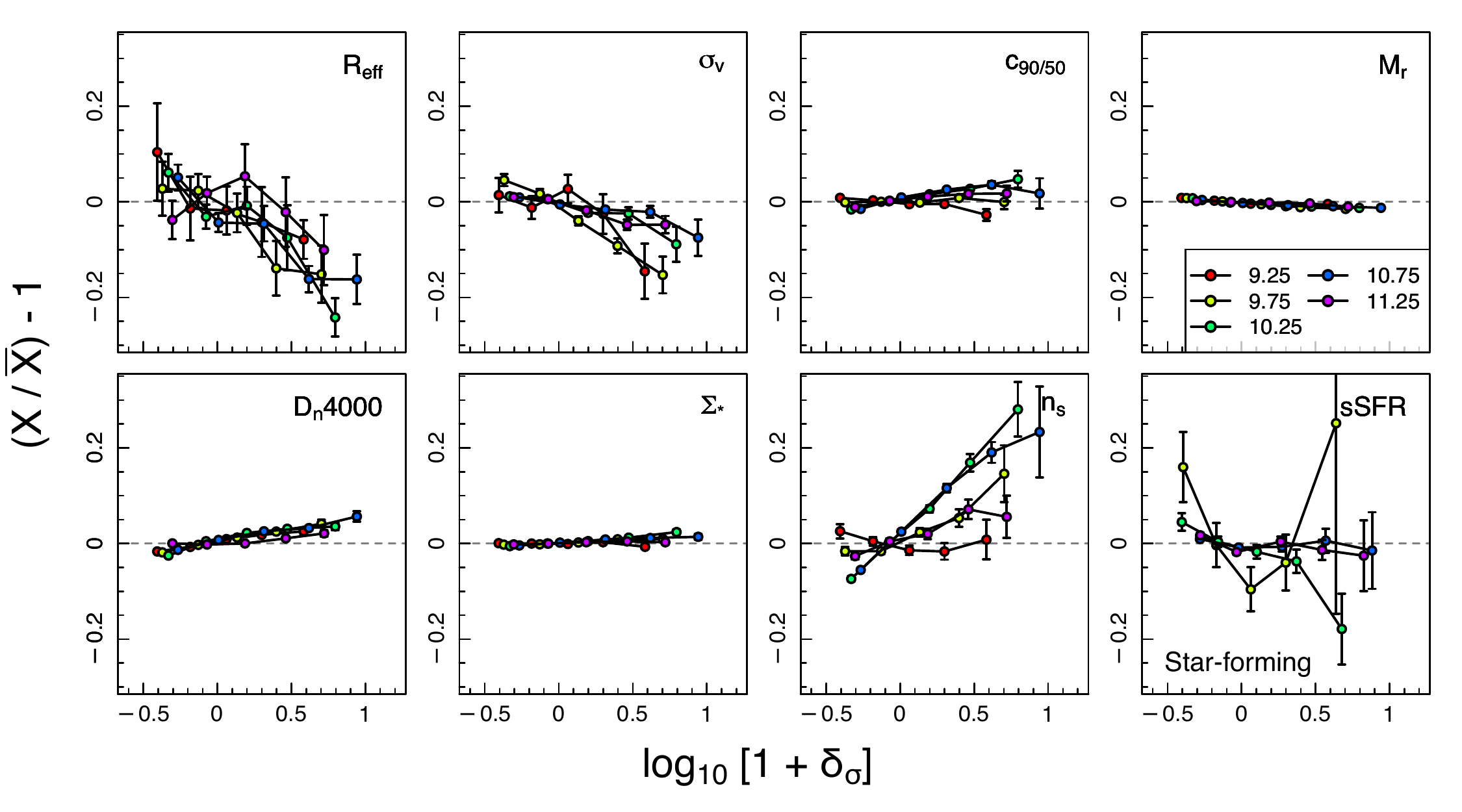}\\
	\includegraphics[width=1.0\textwidth]{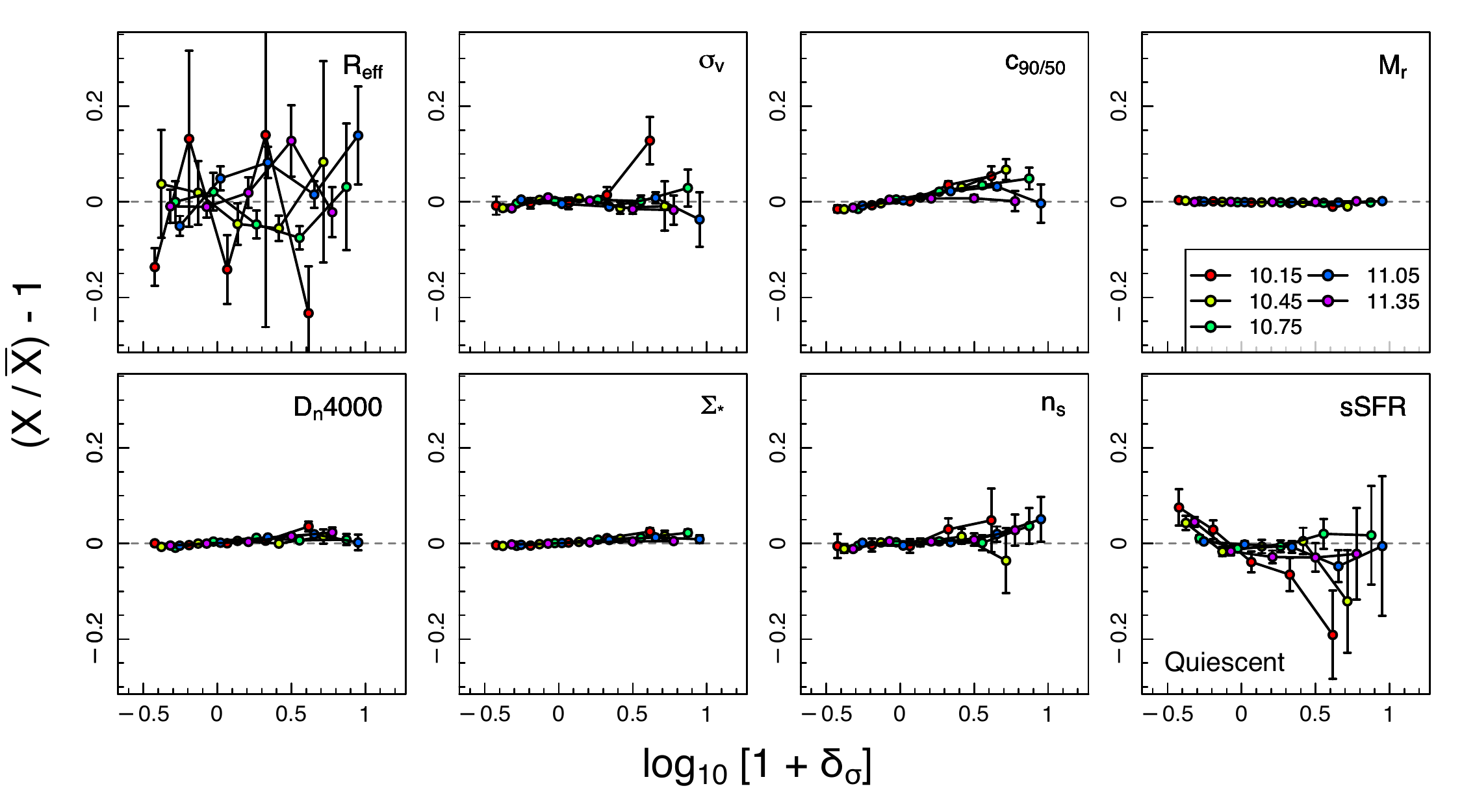}
	\caption{Galaxy parameters plotted as a function of $\log_{10} (1+\delta_{\sigma})$, their normalized number density within 10 Mpc. As with Figure \ref{fig:lsatdens}, we plot along the y-axis the ratio between each galaxy property and a power law fit to that galaxy property in each mass bin. Each panel displays the distribution for a different parameter, denoted in the upper right; and each line corresponds to a different bin of fixed stellar mass. Errors are derived by 100 bootstrap resamples. The top panel displays galaxies with D$_n4000\leq 1.6$ (blue, or star-forming) and the bottom panel galaxies with D$_n4000\geq 1.6$ (red, or quiescent). Note the difference in mass ranges between the two figures; due to the smaller number of red galaxies below $M_* = 10^{10}$ M$_{\odot}$, we only include galaxies above this mass threshold when making this figure. We also do not show the lowest mass bin for star-forming galaxies in the sSFR panel due to the extremely noisy nature of this data.}
	\label{fig:densgrid}
\end{figure*}

To investigate these effects, we plot secondary central galaxy properties against $\log_{10} (1+\delta_{\sigma)}$ in bins of fixed stellar mass in Figure \ref{fig:densgrid}. As with Figure \ref{fig:lsatdens} we plot each galaxy property as the ratio between the measured value and the power law fit to that galaxy property and stellar mass; this ensures our results are free from stellar mass bias. In this figure the top plot shows this relationship for galaxies with D$_n4000\geq 1.6$ and the bottom plot for those with D$_n4000\leq 1.6$. Splitting galaxies by their 4000 \r{A}ngstrom break in this manner provides a clean sample of quiescent and star-forming sub-populations of galaxies (more so than doing a split by colour, as D$_n4000$ is less sensitive to dust) and can also be thought of splitting galaxies according to starformers and quiescent galaxies. Throughout this paper, in the text and figures, we refer to galaxies with D$_n4000\geq 1.6$ as `red' or `quiescent;' and those with D$_n4000\leq 1.6$ as `blue' or `star-forming.' In each plot, each panel shows one of the eight galaxy parameters discussed in this section, normalized by the mean value of that parameter in that mass bin. We note that for quiescent galaxies, no properties correlate with $\delta_\sigma$ with the exception of a weak trend with sSFR. For star-forming galaxies with D$_n4000\leq 1.6$ we see trends in $\reff$ and $\sigma_v$ where we see larger galaxies with lower velocity dispersions in more dense environments. We also see a trend in $n_s$ for star-forming galaxies, where it is flat at low M$_*$ and rising at high M$_*$. Such a trend is indicative of a correlation with halo mass: at fixed M$_*$ the distribution of M$_h$ gets very broad at high M$_*$, and can manifest as a correlation with density \citep{Wang2018}. Recent work by \citep{Calderon2018} detects a tentative signal for 2-halo conformity when examining the S\'ersic indices of central galaxies, suggesting that galaxy morphology may be a strong tracer of halo assembly bias. Finally, we also see a small inverse relationship between density and specific star-formation rate for star-forming galaxies, which is consistent with recent results from \citet{Tinker2017}. We note here that there are no significant changes in the relationship between density and $\lsat$ shown in Figure \ref{fig:lsatdens} when we split by D$_n4000$. 

\subsection{Central galaxy properties as a function of $\lsat$}

Figures \ref{fig:LsatParms1} and \ref{fig:LsatParms2} show the relationship between $\lsat$ and the 8 aforementioned galaxy properties. All figures have the same format, consisting of two panels. The left panel will show mean $\lsat$ values for quantiles of the parameter of interest in bins of fixed stellar mass (e.g. the mean $\lsat$ value of galaxies whose effective radius is $ \geq 90^{\mathrm{th}}$ percentile of galaxies with $10 \leq \log_{10} \msun < 10.1$). The right hand panel will show the same data, but this time with mean values of $\lsat$ shown at fixed values of the parameter in question in different stellar mass bins. Stellar mass values shown in the legend of each figure are the mid-points of that mass bin. In all cases, errors are computed using 100 bootstrap resamples of the data.

\subsubsection{Effective radius}

\begin{figure*}
	\centering
	\includegraphics[width=0.495\textwidth]{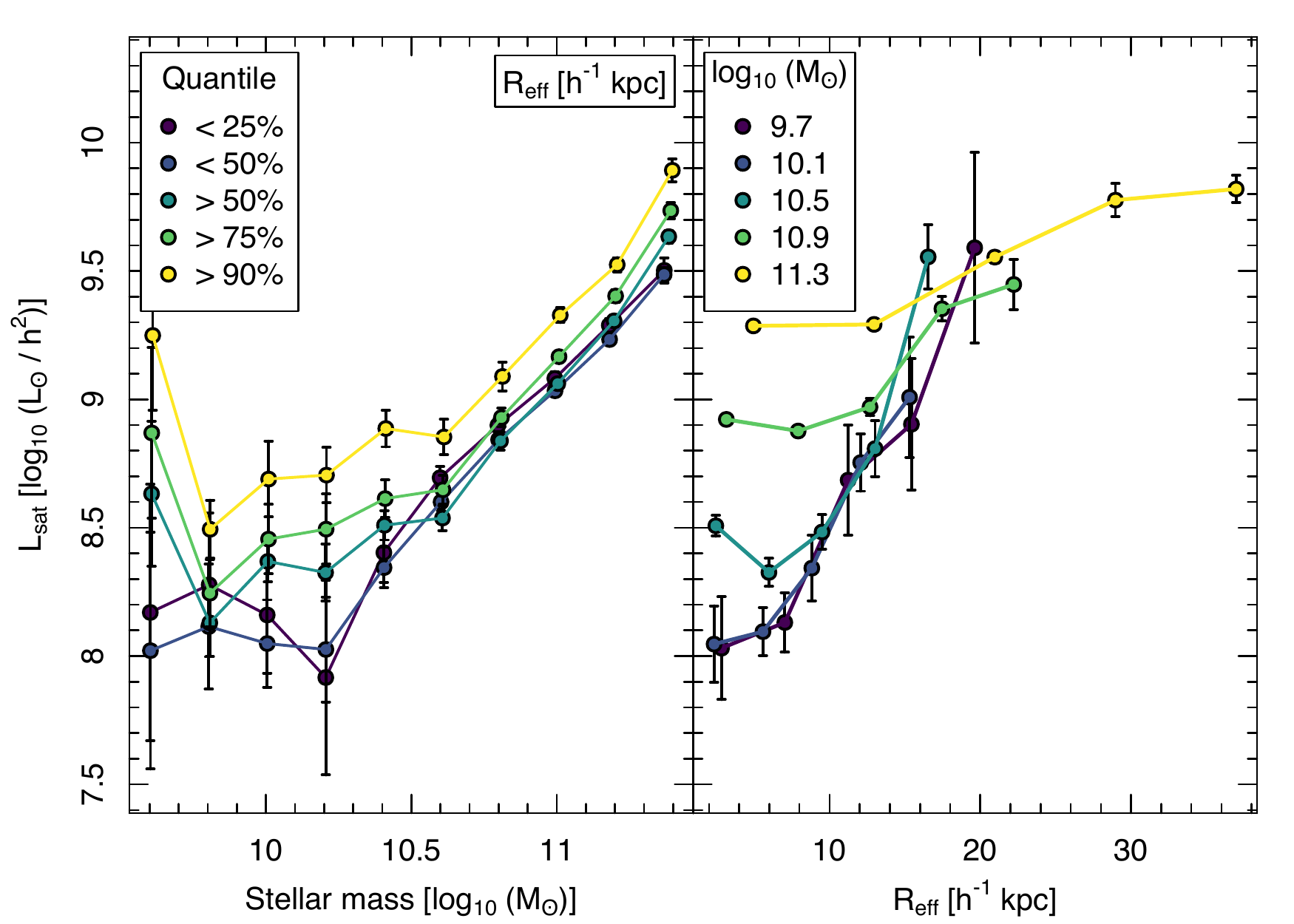} 
	\includegraphics[width=0.495\textwidth]{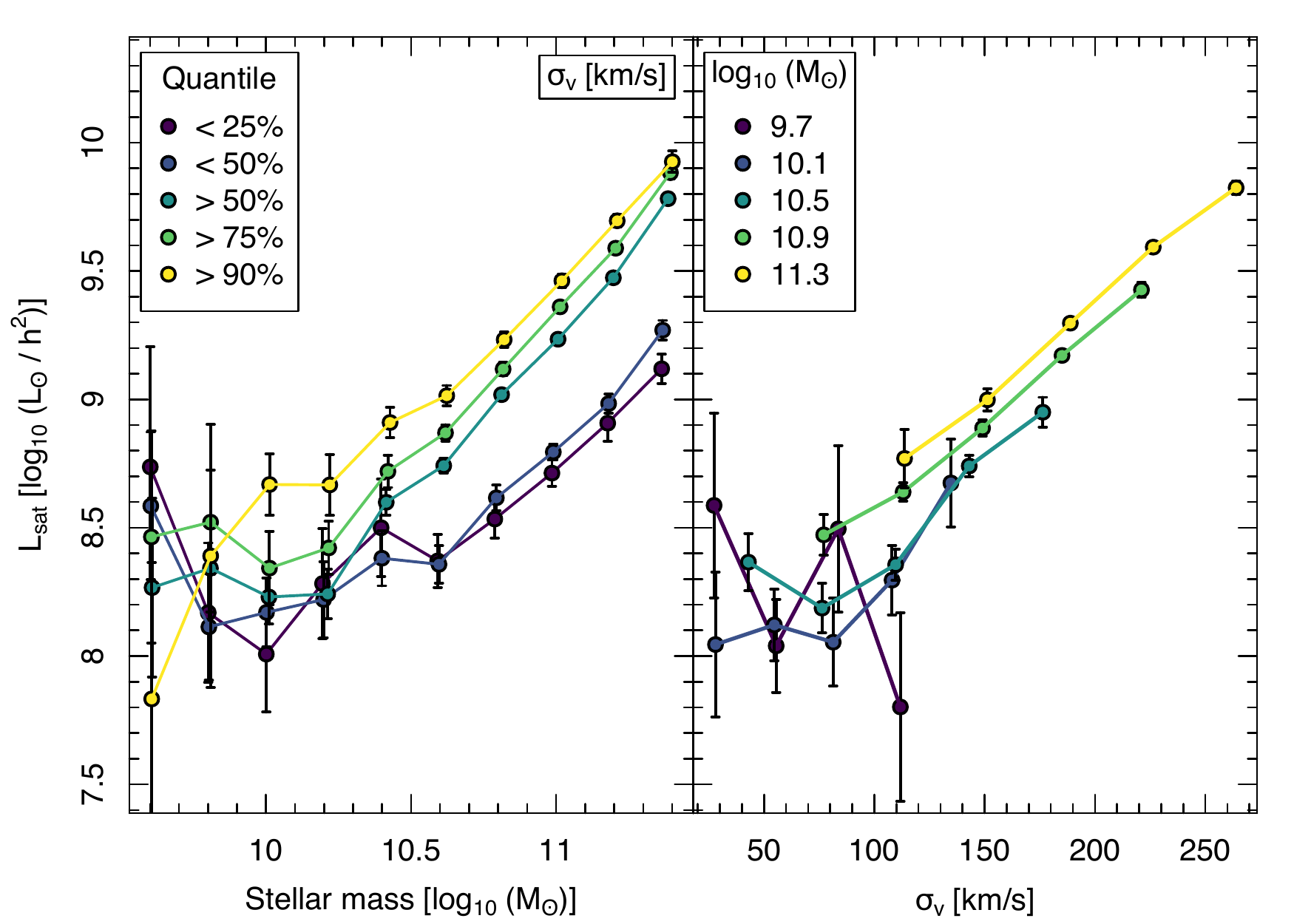} 
	\includegraphics[width=0.495\textwidth]{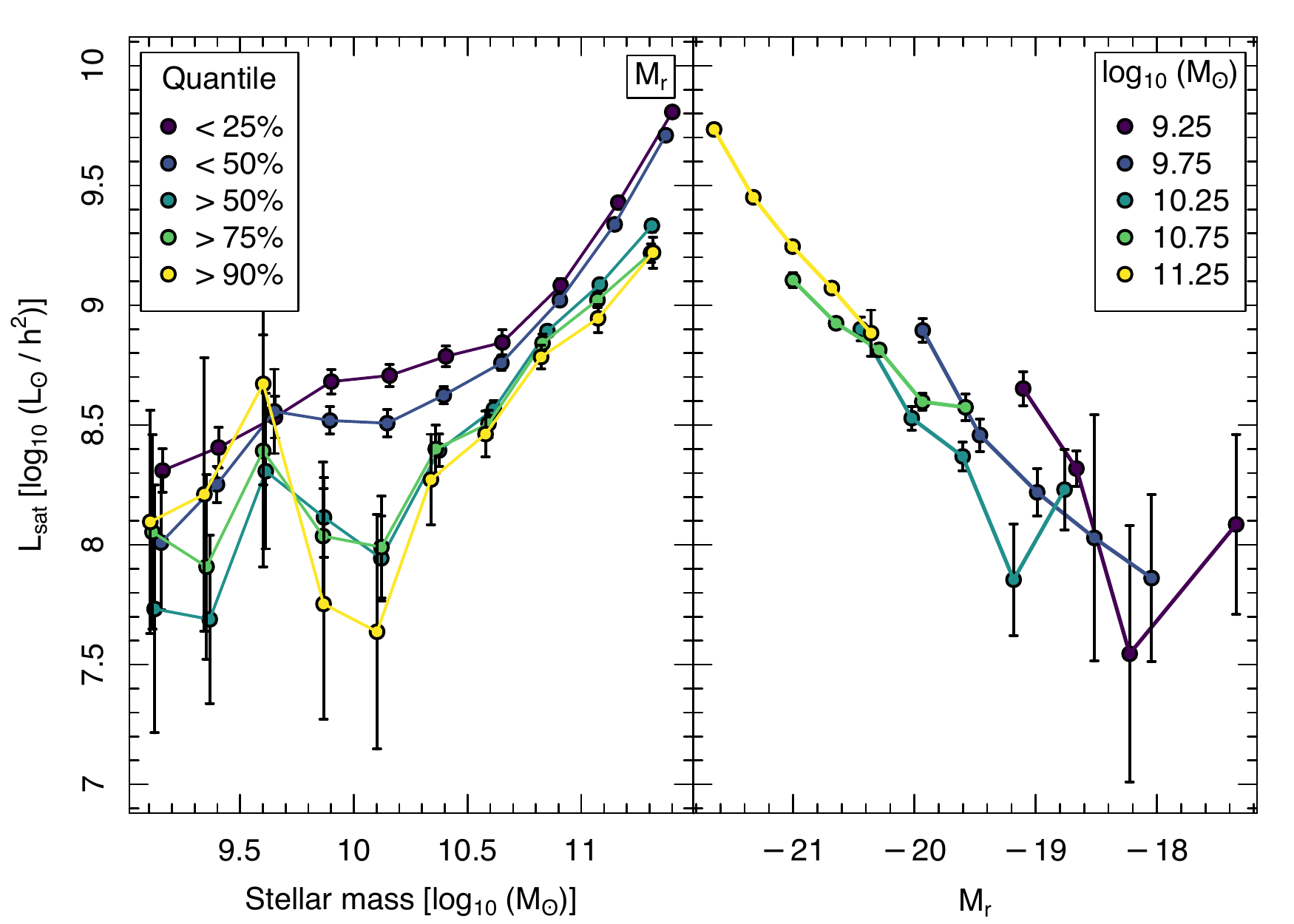}
	\includegraphics[width=0.495\textwidth]{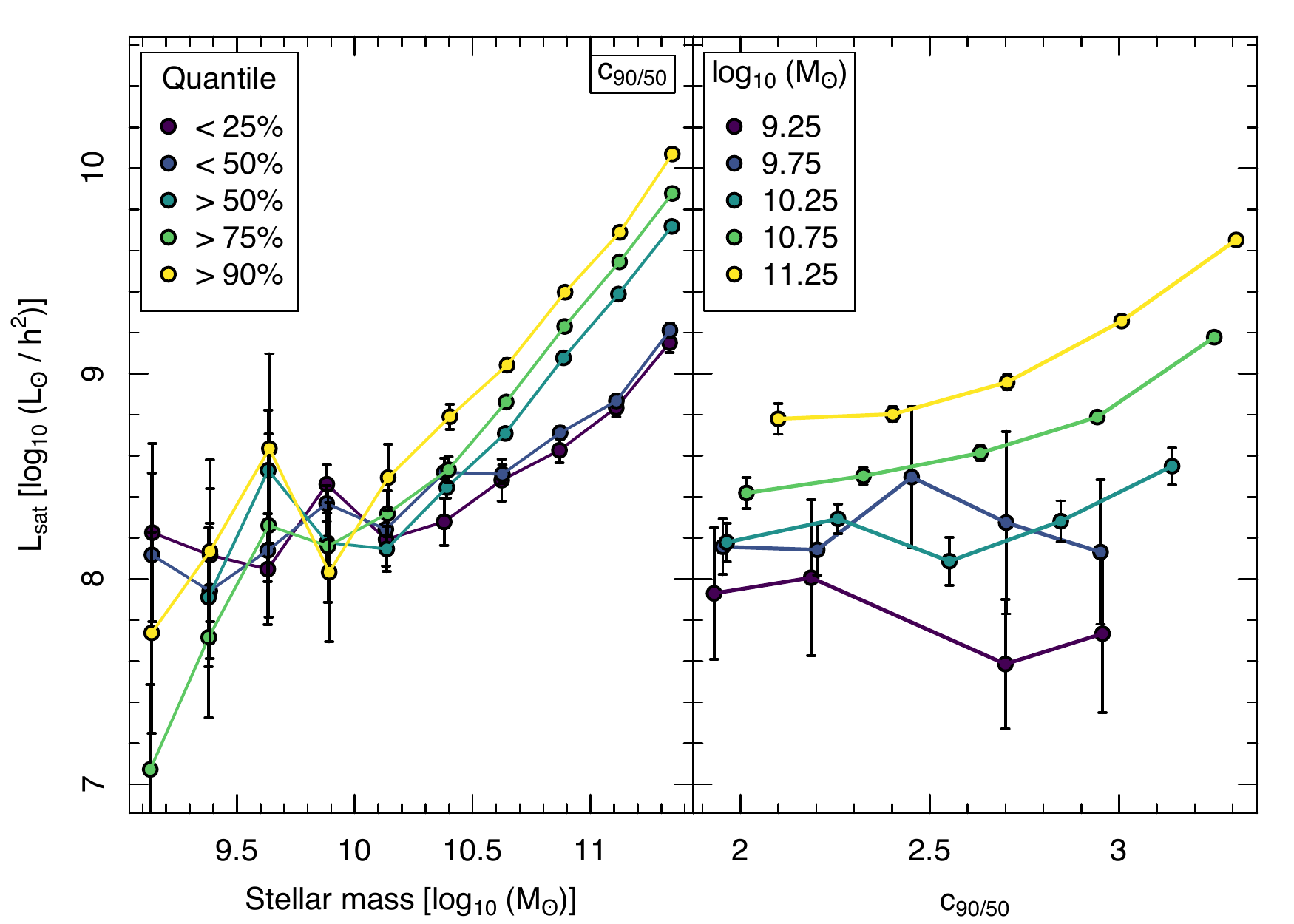}
	\caption{Relationships between $\lsat$ and $\reff$ (top-left), $\sigma_v$ (top-right), $M_r$ (bottom-left), and concentration (bottom-right). In the left-hand panel of each figure we plot $\lsat$ in bins of fixed stellar mass. In each stellar mass bin we identify quantiles of the galaxy parameter in question and plot the mean $\lsat$ value of galaxies in those quantiles, each of which is shown as a separate set of points. On the right-hand panel of each figure we show $\lsat$ in bins of fixed parameter space in different intervals of stellar mass, each of which is shown by a different set of points. In all cases errors shown are standard errors estimated using 100 bootstrap resamples.}
	\label{fig:LsatParms1}
\end{figure*}

\begin{figure*}
\includegraphics[width=0.495\textwidth]{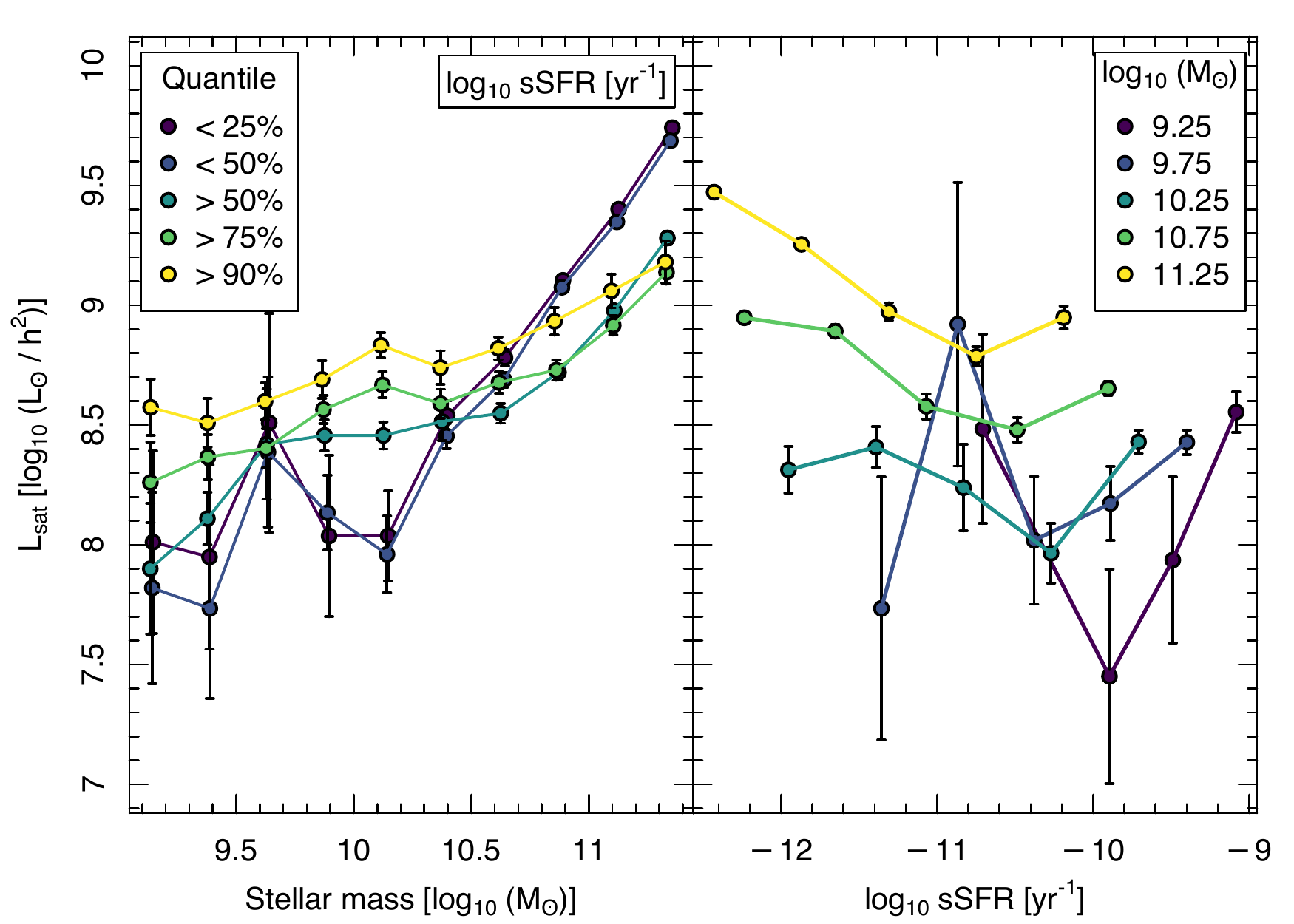}
	\includegraphics[width=0.495\textwidth]{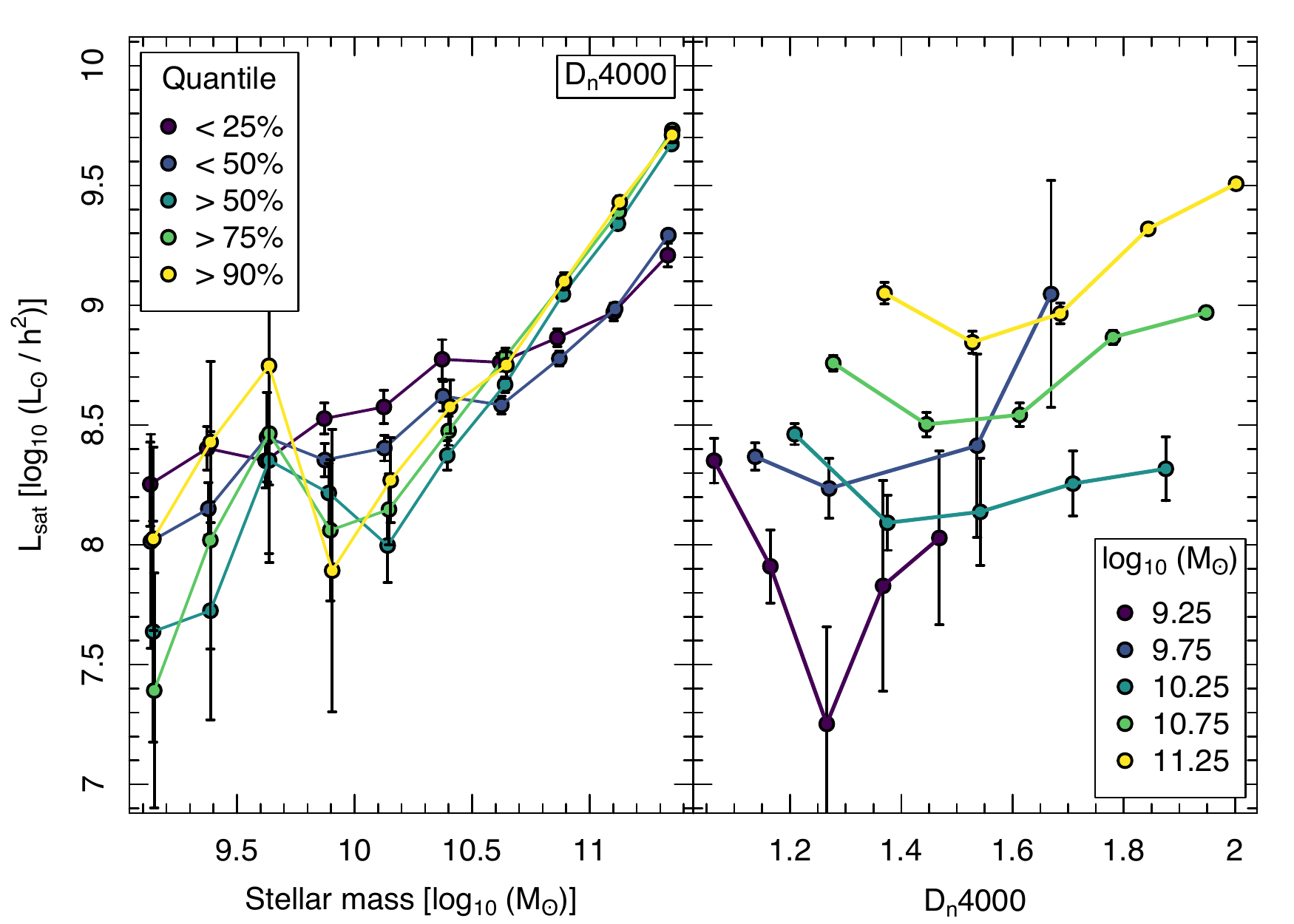}
	\includegraphics[width=0.495\textwidth]{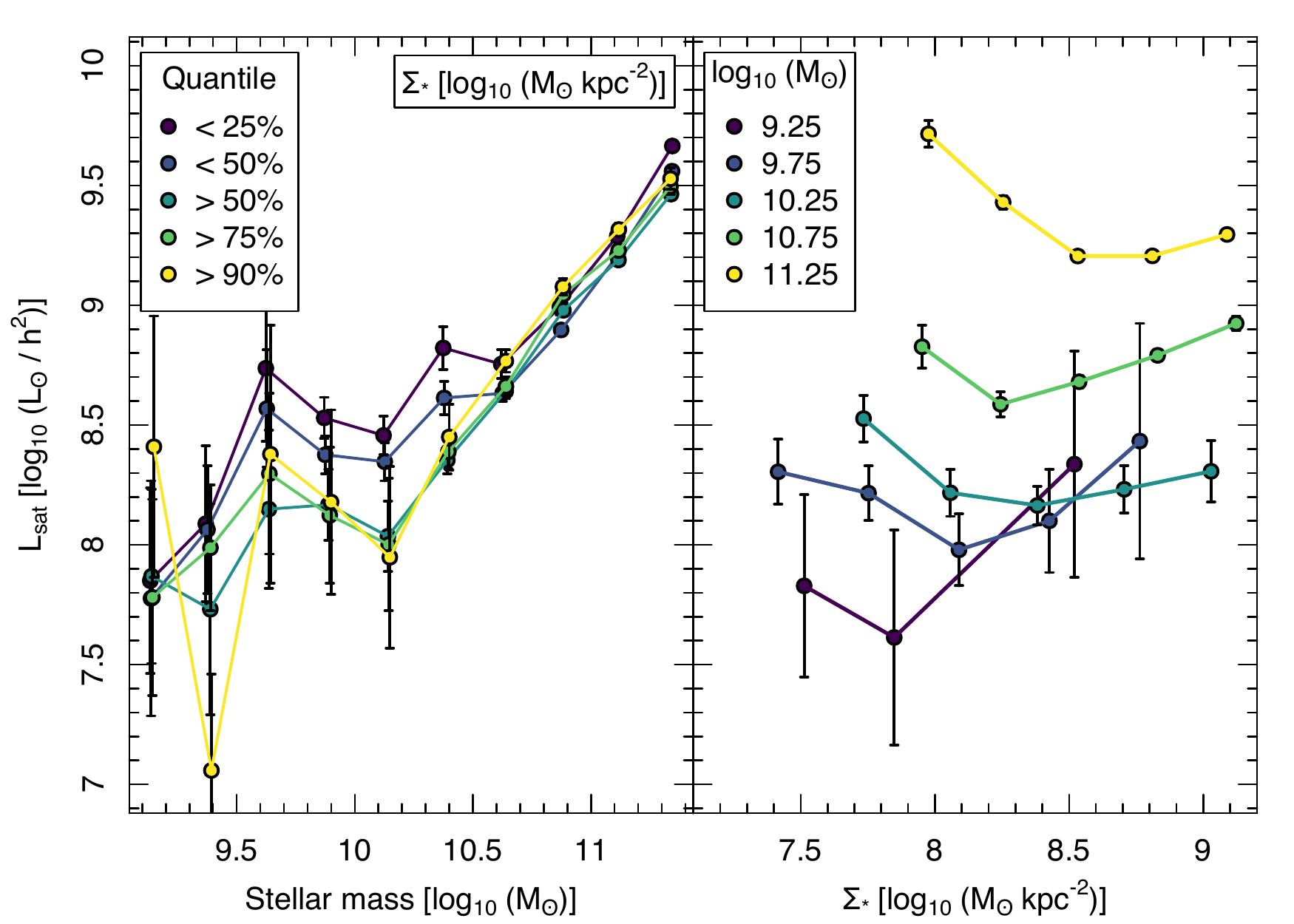}
	\includegraphics[width=0.495\textwidth]{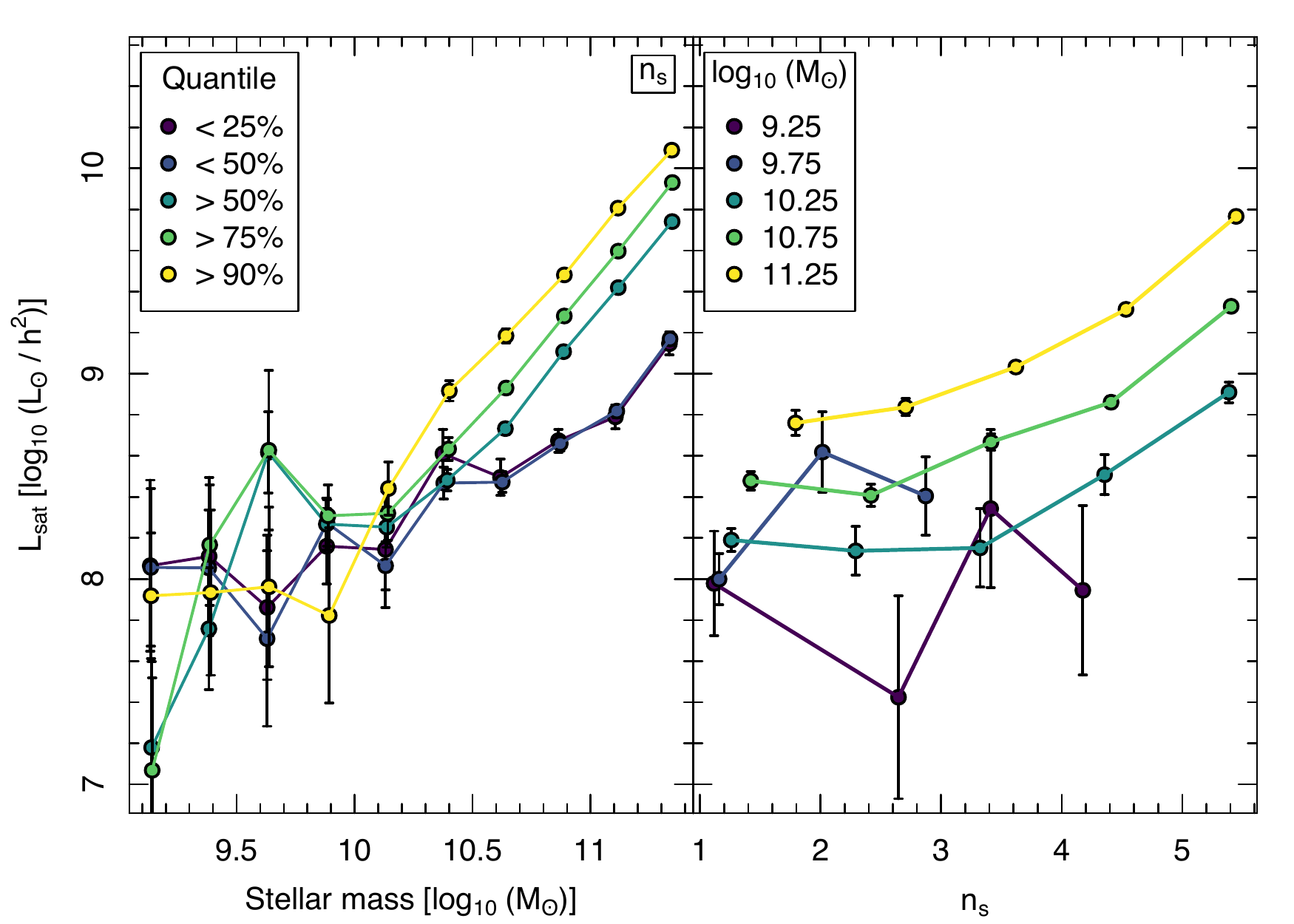}
	\caption{Relationships between $\lsat$ and sSFR (top-left), D$_n4000$ (top-right), $\Sigma_*$ (bottom-left), and S\'ersic index (bottom-right). In the left-hand panel of each figure we plot $\lsat$ in bins of fixed stellar mass. In each stellar mass bin we identify quantiles of the galaxy parameter in question and plot the mean $\lsat$ value of galaxies in those quantiles, each of which is shown as a separate set of points. On the right-hand panel of each figure we show $\lsat$ in bins of fixed parameter space in different intervals of stellar mass, each of which is shown by a different set of points. In all cases errors shown are standard errors estimated using 100 bootstrap resamples.}
	\label{fig:LsatParms2}
\end{figure*}

We utilize the \texttt{SERSIC\_TH50} parameter from the NSA catalogue to represent effective radius. \texttt{SERSIC\_TH50} measures the radius along the major axis of the 2D S\'ersic fit to the galaxy. The top-left panel of Figure \ref{fig:LsatParms1} displays the relationship between effective radius and $\lsat$ as outlined in the preceding section. Below stellar masses of order $10^{10.25} \msun$, $\reff$ is a significantly better predictor of $\lsat$ than stellar mass. At higher masses the effective radius performs well at predicting $\lsat$ compared to stellar mass.     

\subsubsection{Velocity dispersion}

The velocity dispersion of each galaxy is taken from the NYU-VAGC catalog, and is given in kilometres per second. These dispersions are measured directly from the spectra. The top-right panel of Figure \ref{fig:LsatParms1} shows the relationship between $\lsat$ and $\sigma_v$ as a function of stellar mass. From the right hand panel of this figure it can be seen that past a stellar mass threshold of approximately $10^{10.5}$ $\msun$, $\sigma_v$ performs better as a predictor of $\lsat$ than stellar mass. Even at stellar masses $> 10^{10}$ $\msun$, separating the galaxy population into rank-ordered quantiles of $\sigma_v$ shows that this parameter is a power discriminant of $\lsat$. 

\subsubsection{Absolute r-band magnitude}

The absolute r-band magnitude $M_{\mathrm{r}}$ of a central SDSS galaxy shows a strong correlation to $\lsat$ that is even stronger than its correlation with stellar mass, as shown in the right hand panel of the bottom-left panel of Figure \ref{fig:LsatParms1}. These values are given under the \texttt{ABSMAG} column of the NYU-VAGC catalogue. As before, the trends are stronger at stellar masses greater than $10^{10}$ $\msun$, which is also seen in the left hand panel (note the reversed trend in the quantiles for stellar mass here, due to smaller magnitudes corresponding to brighter objects).

\subsubsection{Concentration}

Each galaxy's concentration, c$_{90/50}$, is computed by taking the ratio between $r_{50}$ and $r_{90}$, the radii that enclose $50\%$ and $90\%$ of the total integrated light from the galaxy respectively. This parameter is given in the NYU-VAGC catalogue. The relationship betwen $\lsat$ and concentration does not show as striking a correlation as the previous parameters we have already examined, particularly at lower stellar masses, as can be seen in the right hand panel of the bottom-right plot in Figure \ref{fig:LsatParms1}. Nevertheless, at stellar masses greater than $10^{10.5}$ h$^{-1}$ Mpc, concentration still shows strong trends with $\lsat$.

\subsubsection{Star formation indicators}

The top-left and top-right panels of Figure \ref{fig:LsatParms2} show the relationship between $\lsat$ and two indicators of star formation measured from \r{s}pectra; namely the 4000 \r{A}ngstrom break (D$_n4000$, taken from the NYU-VAGC catalog) and the specific star formation rate of each galaxy, which is computed from nebular emission lines and galaxy photometry (we specifically use the \texttt{SPECSFR\_TOT\_P50} parameter) as computed by \citet{Brinchmann2004} and given in the MPA-JHU catalogue. As with the other parameters shown in this section, both D$_n4000$ and sSFR correlate well with $\lsat$, though the correlation is stronger at higher masses than for other parameters shown before. Part of this trend is caused by the fact that we are including both active and passive galaxies in our sample, which will have the effect of adding extra noise to $\lsat$. The D$_n4000$ - $\lsat$ relationship is further complicated by the bimodal distribution of this parameter. 

\subsubsection{Mass surface density}

Mass surface densities are computed by dividing the stellar mass of the galaxy by its effective radius, and are given in the NSA catalogue. The right hand panel of the bottom-left plot in Figure \ref{fig:LsatParms2} suggests that this parameter does not show such a strong correlation with $\lsat$ as we have seen in previous sections. Galaxies with high stellar masses and low mass surface densities do show some trends with $\lsat$, but by and large this parameter is not a strong discriminant of total satellite luminosity. This is likely caused by the degeneracy between stellar mass and mass surface density. We explore this relationship further in Section \ref{sec:multi}.

\subsubsection{S\'ersic index}

S\'ersic indices are computed from two-dimensional fits to the surface brightness profiles of each galaxy, and is provided as \texttt{SERSIC\_N} in the NSA catalogue. As with parameters like concentration, the S\'ersic index only shows strong correlations with $\lsat$ at stellar masses exceeding $10^{10.5}$ h$^{-1}$ Mpc. This relationship is shown in the bottom-right plot of Figure \ref{fig:LsatParms2}

\subsection{Multivariate relationships with $\lsat$}
\label{sec:multi}

Many of the parameters we have examined in the preceding subsection are correlated with each other (e.g. $c_{90/50}$, $\sigma_v$, and $\reff$). Some of the results shown in Figures \ref{fig:LsatParms1} and \ref{fig:LsatParms2} are seemingly incompatible with each other within the context of how these parameters correlate with each other: specifically, we know that $\reff$ and $\sigma_v$ are anti-correlated in galaxies, yet they both yield strong positive correlations with $\lsat$. Until now we have explored how $\lsat$ varies as a function of one parameter at a time; however, given these observations it can be informative to look at how $\lsat$ behaves with pairs of parameters whose interrelationships are already well understood. This also has the potential of unrooting any possible degeneraces that may be driving the trends seen above. To this end, we now present some figures showing $\lsat$ binned with two parameters at once. As before, these figures will have the same format: a two dimensional histogram of the mean value of $\lsat$ in galaxies binned by two parameters. Bins containing fewer than 10 galaxies are not filled in, and in addition we include contours highlighting the 25th, 50th, 75th, and 90th percentiles of $\lsat$ within the entire distribution. These contours are drawn from the 2D histogram after it has been smoothed by being convolved with a Gaussian whose FWHM is 3 times the pixel size of each figure.

\begin{figure*}
	\centering
	\includegraphics[width=0.495\textwidth]{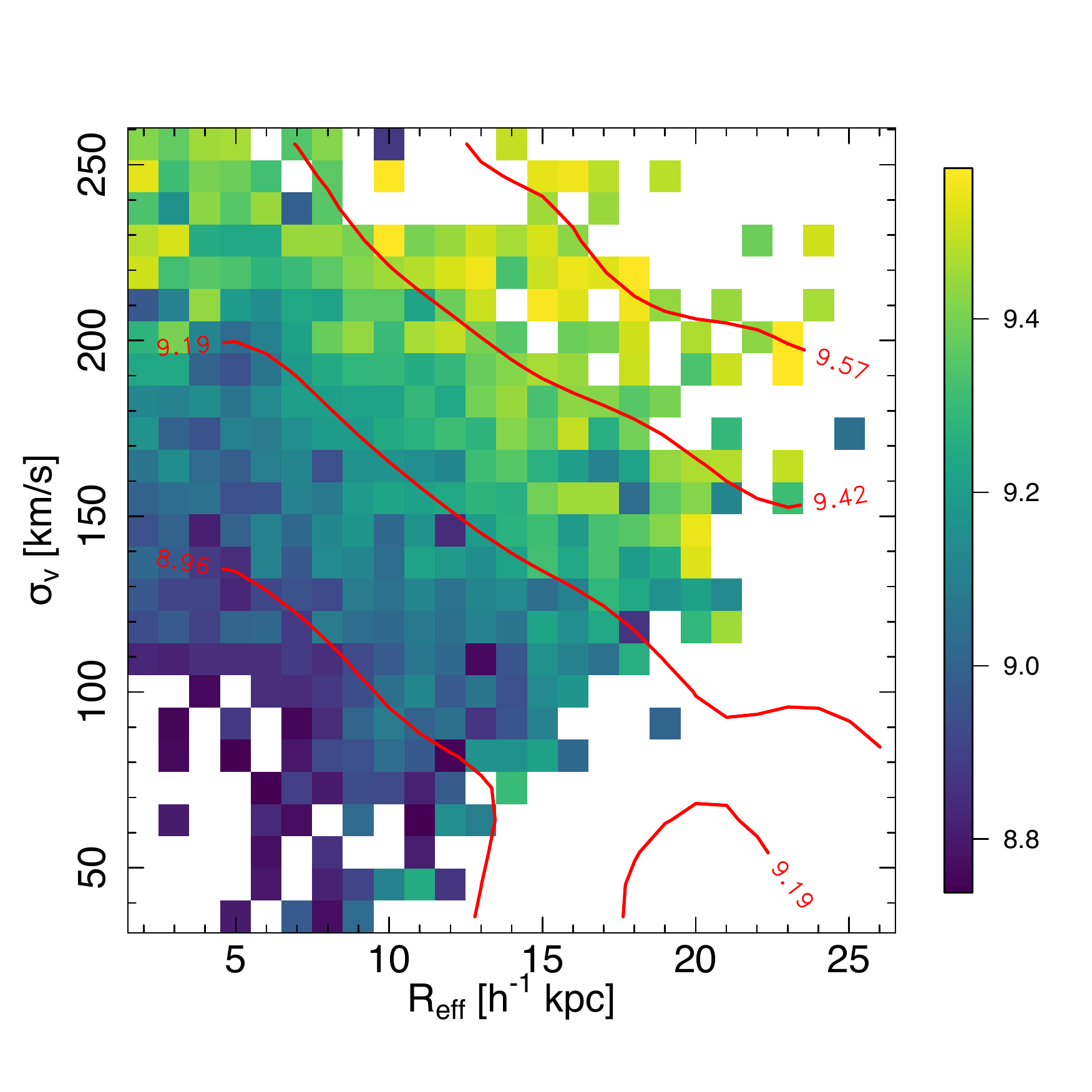} 
	\includegraphics[width=0.495\textwidth]{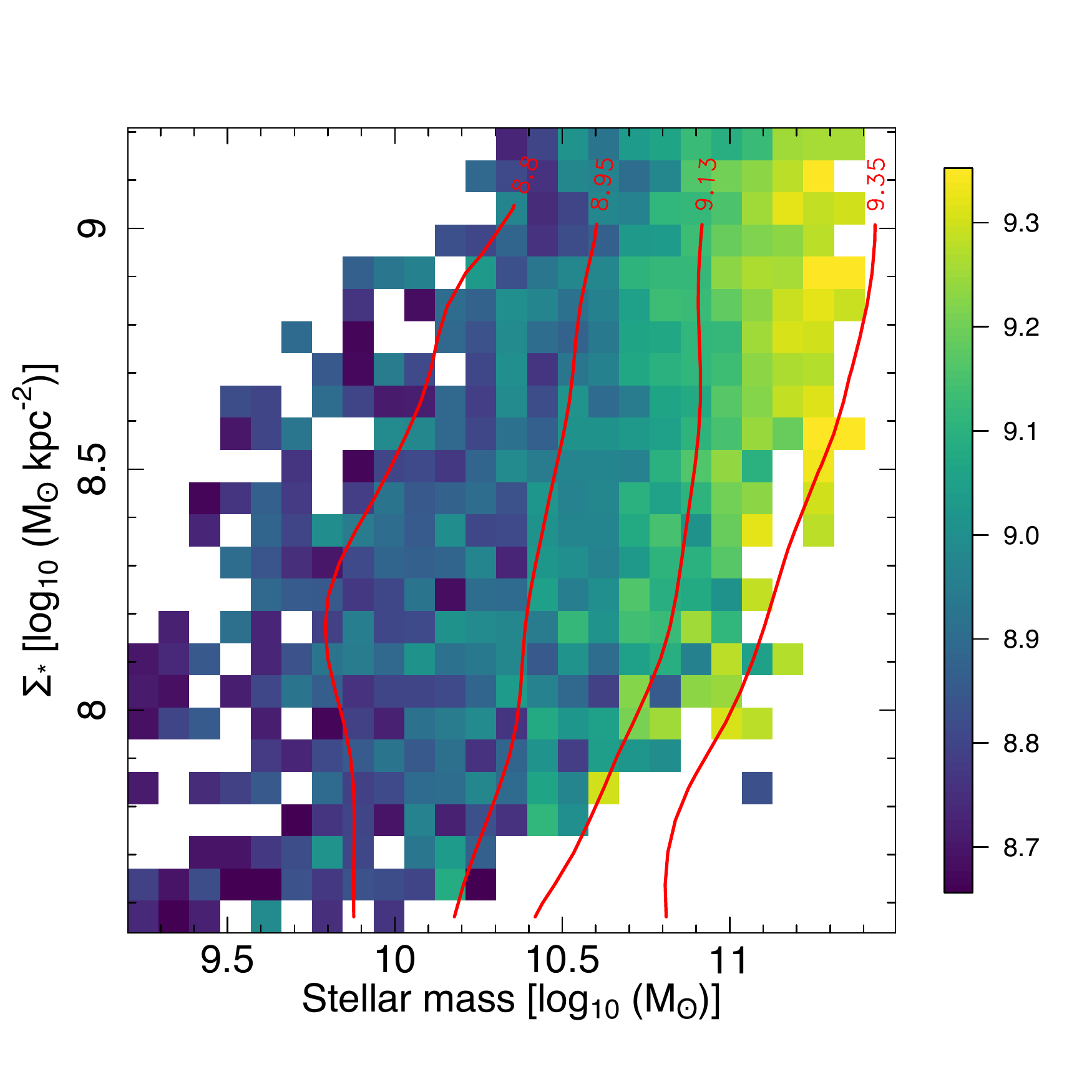}
	\includegraphics[width=0.495\textwidth]{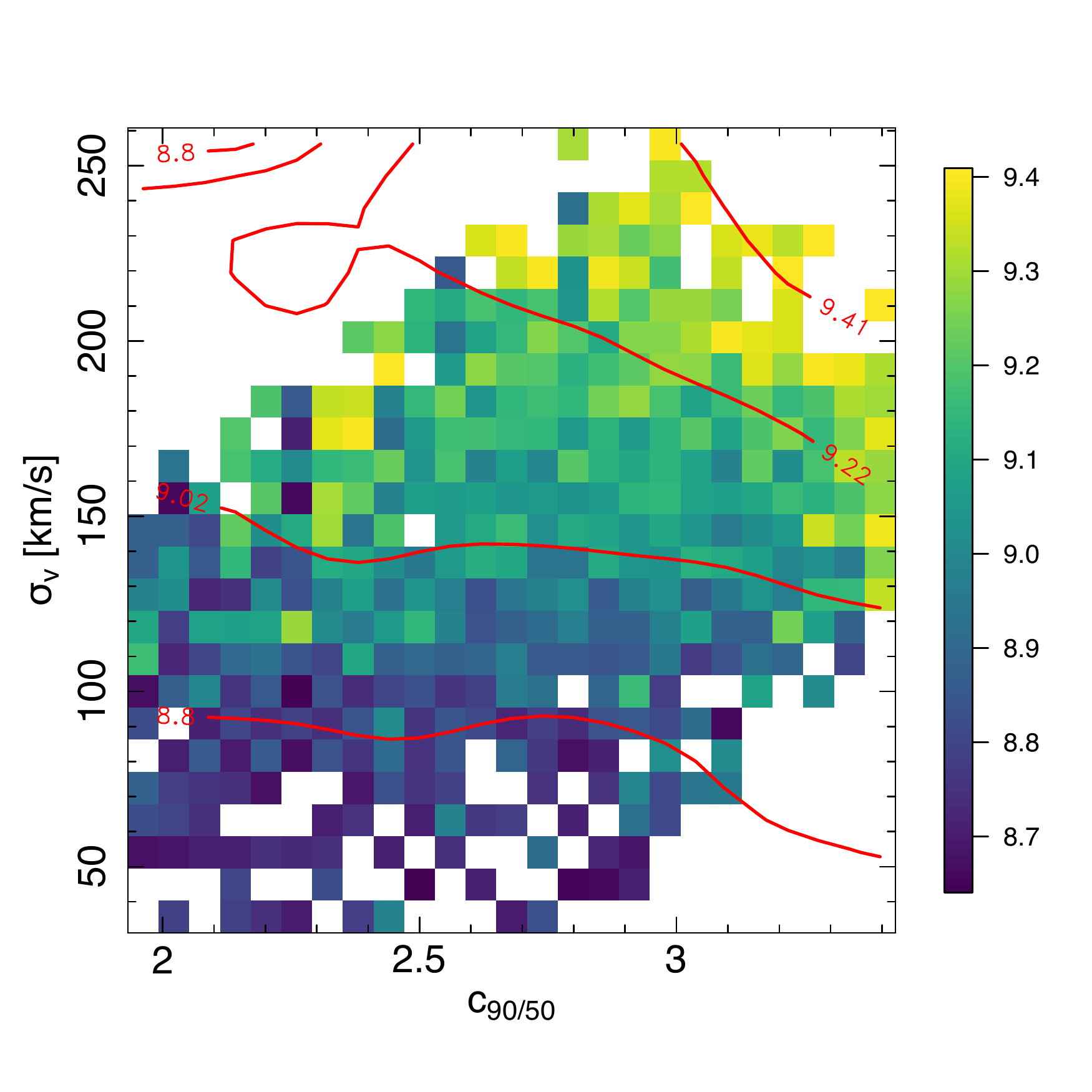}
	\includegraphics[width=0.495\textwidth]{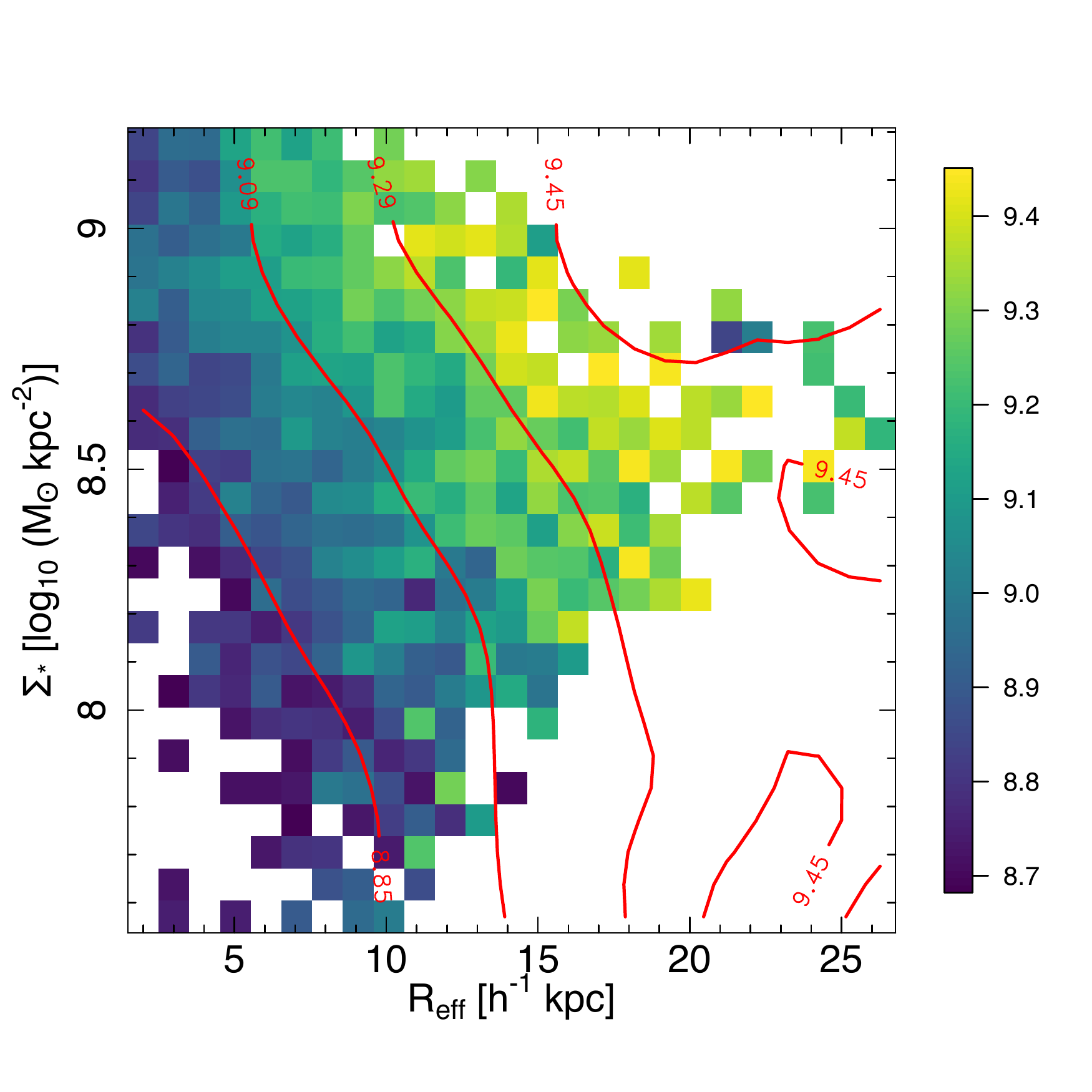}
	\caption{Two-dimensional histograms of $\lsat$ plotted at fixed values of combinations of two parameters. \emph{Top-left:} $\reff$ and $\sigma_v$. \emph{Top-right:} $M_*$ and $\Sigma_*$. \emph{Bottom-left:} Concentration and $\sigma_v$. \emph{Bottom-right:} $\reff$ and $\Sigma_*$. In all panels the color scale represents the logarithm of the mean value of $\lsat$ in that bin. Bins with fewer than 10 galaxies are not shown. Contour lines mark the 25th, 50th, 75th, and 90th percentiles of $\lsat$ within the entire distribution and are drawn over a smoothed version of the grid. Smoothing is done by convolving the 2D histogram with a Gaussian kernel whose full-width at half maximum is 3$\times$ the pixel size.}
	\label{fig:lsatgrids}
\end{figure*}

In Figure \ref{fig:lsatgrids} we display trends in $\lsat$ at fixed values of $\reff$ and $\sigma_v$ (top-left panel); $M_*$ and $\Sigma_*$ (top-right panel); concentration and $\sigma_v$ (bottom-left panel); and $\reff$ and $\Sigma_*$ (bottom-right panel). We see from the top two panels that $\lsat$ behaves `as expected' within these parameter spaces: as both $\reff$ and $\sigma_v$ increase, we see an increase in $\lsat$ that is orthogonal to these parameters, consistent with a picture in which large galaxies with complex stellar orbits (i.e. galaxies that have undergone many mergers) play host to a greater number of satellites. By contrast, at fixed stellar mass $\Sigma_*$ plays a minimal role in determining $\lsat$, which is consistent with the left hand panel of the bottom-left plot in Figure \ref{fig:LsatParms2}.

We can explore this relationship further by looking into how $\lsat$ varies in bins of fixed $\Sigma_*$ and $\reff$, shown in the bottom-left plot of Figure \ref{fig:lsatgrids}. Combined with the results shown in the bottom-right plot of this figure, it can be seen that the sensitivity of $\lsat$ to $\Sigma_*$ at fixed stellar mass is likely driven by the sensitivty of $\lsat$ to $\reff$, and the way in which this latter parameter is strongly correlated with $\lsat$. In other words, within the context of these three parameters, the galaxy parameter that is most likely to impact the total satellite luminosity at fixed stellar mass is its effective radius; and the inter-relationship between size and stellar mass that is used to derive $\Sigma_*$ is what drives the weaker trend with this parameter and $\lsat$.

A notable and interesting feature of the plots shown in Figure \ref{fig:lsatgrids} is that $\lsat$ appears to be more sensitive to certain galaxy properties than others; this is best shown by the slope of the contour lines shown in this Figure. For example, in the bottom-left plot, the contour lines are largely horizontal, showing that $\sigma_v$ correlates more strongly with $\lsat$ than $c_{90/50}$. Similarly, in the bottom-right plot we see that $\reff$ correlates more strongly with $\lsat$ than $\Sigma_v$ from the mostly vertical contours. Meanwhile, in the upper-left plot where we show both $\sigma_v$ and $\reff$ the contours are orthogonal to both axes, suggesting that both play an important role in determining $\lsat$. All of this suggests a hierarchy in the amount of information that galaxy observables provide on $\lsat$. We quantify this in the next section.

\subsection{Quiescent and star-forming galaxies}

\begin{figure*}
	\centering
	\includegraphics[width=0.495\textwidth]{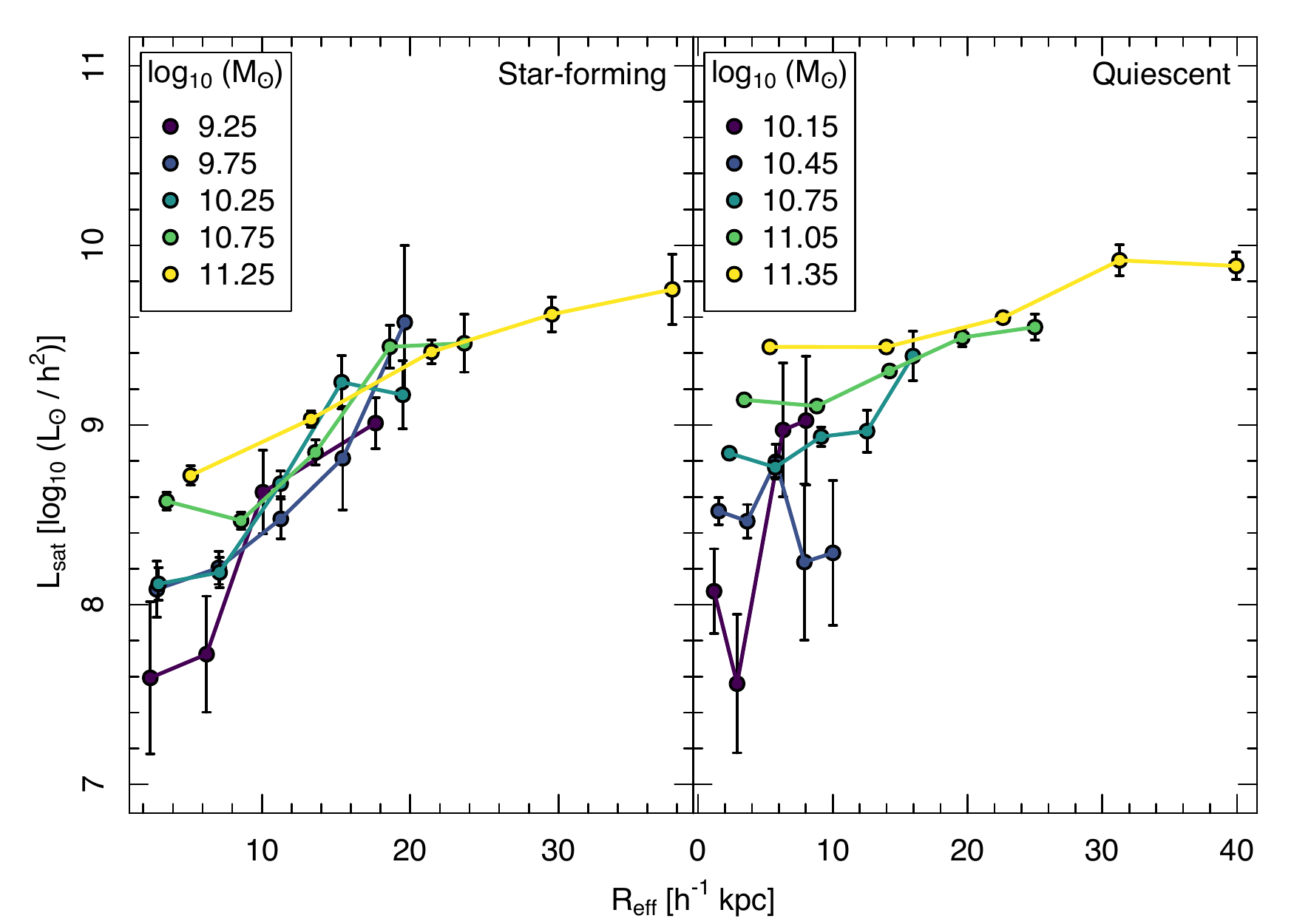} 
	\includegraphics[width=0.495\textwidth]{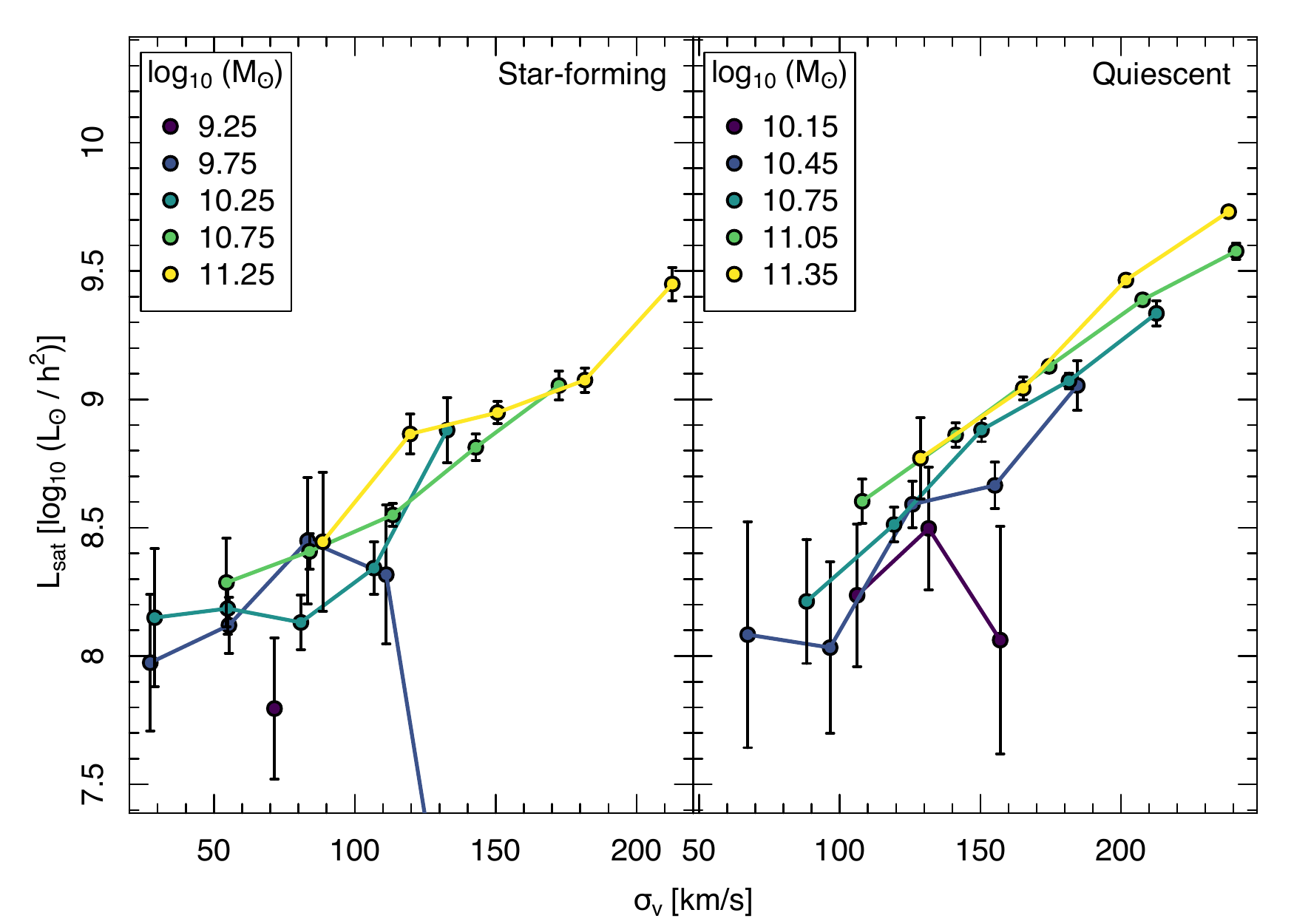} 
	\includegraphics[width=0.495\textwidth]{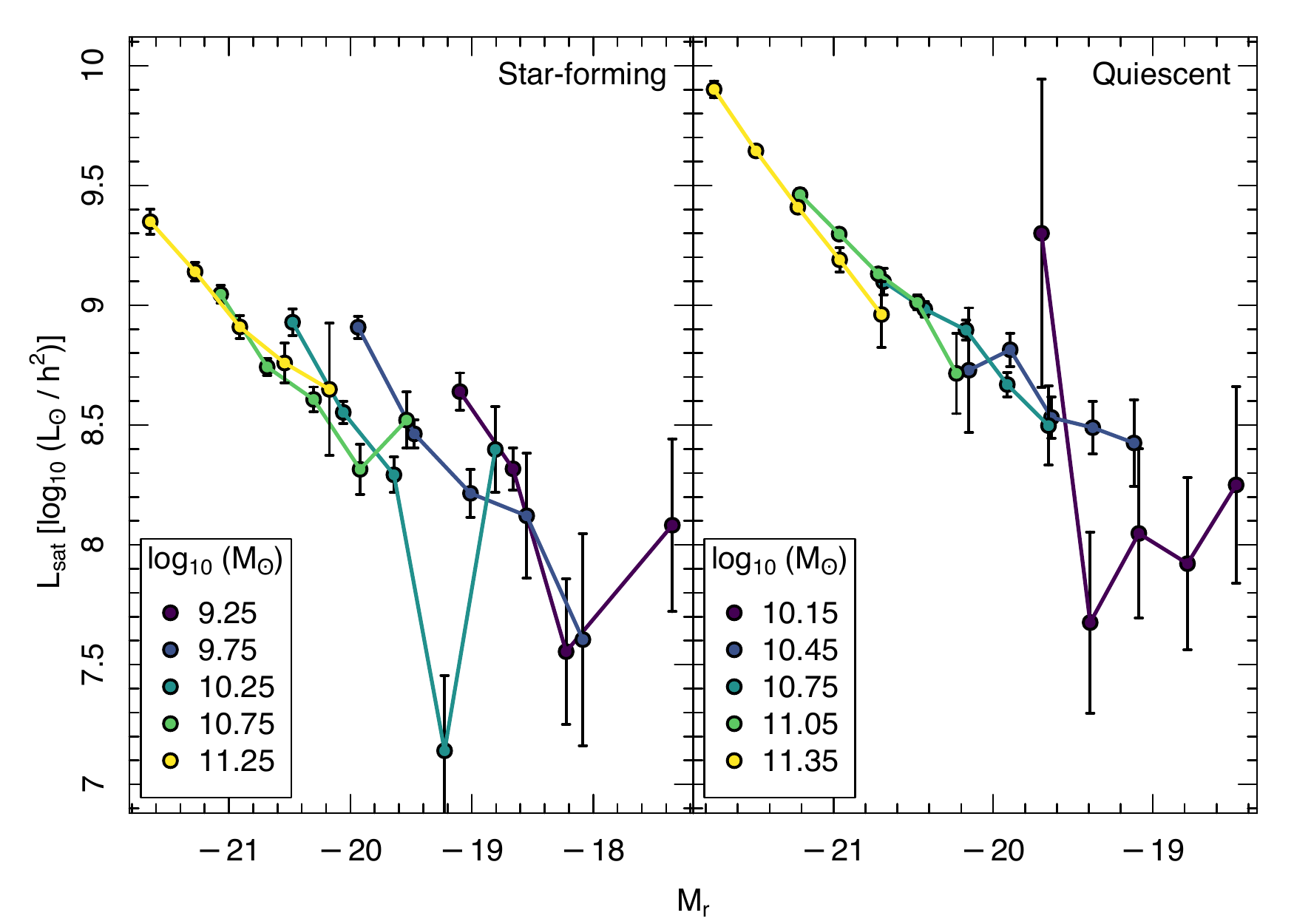}
	\includegraphics[width=0.495\textwidth]{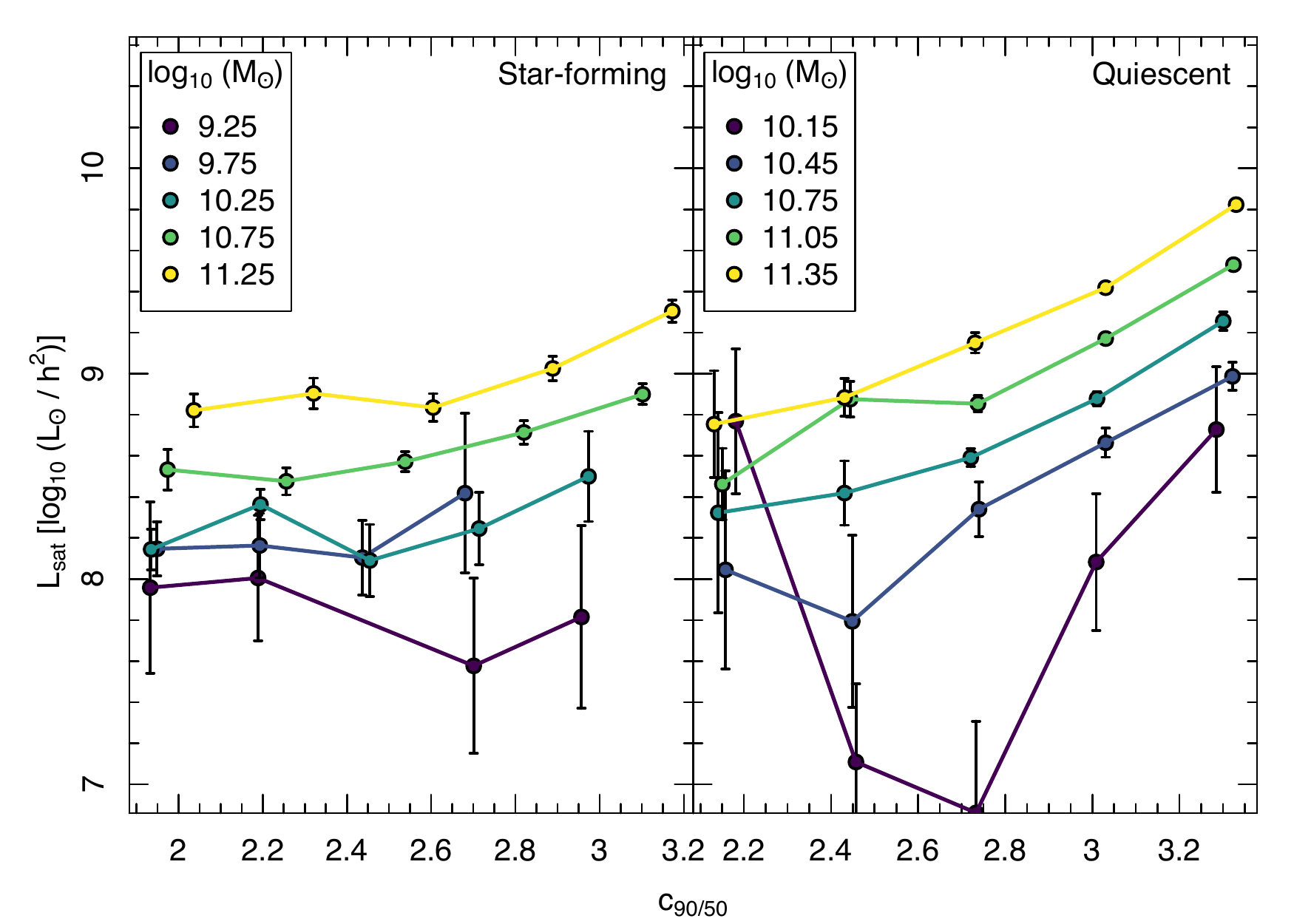}
	\caption{$\lsat$ plotted as a function of $\reff$ (top-left), $\sigma_v$ (top-right), $M_r$ (bottom-left), and concentration (bottom-right) in bins of fixed stellar mass for populations of star-forming and quiescent galaxies, as selected by their D$_n4000$ strength (left and right panels of each plot, respectively).}
	\label{fig:LsatRvB1}
\end{figure*}

\begin{figure*}
\includegraphics[width=0.495\textwidth]{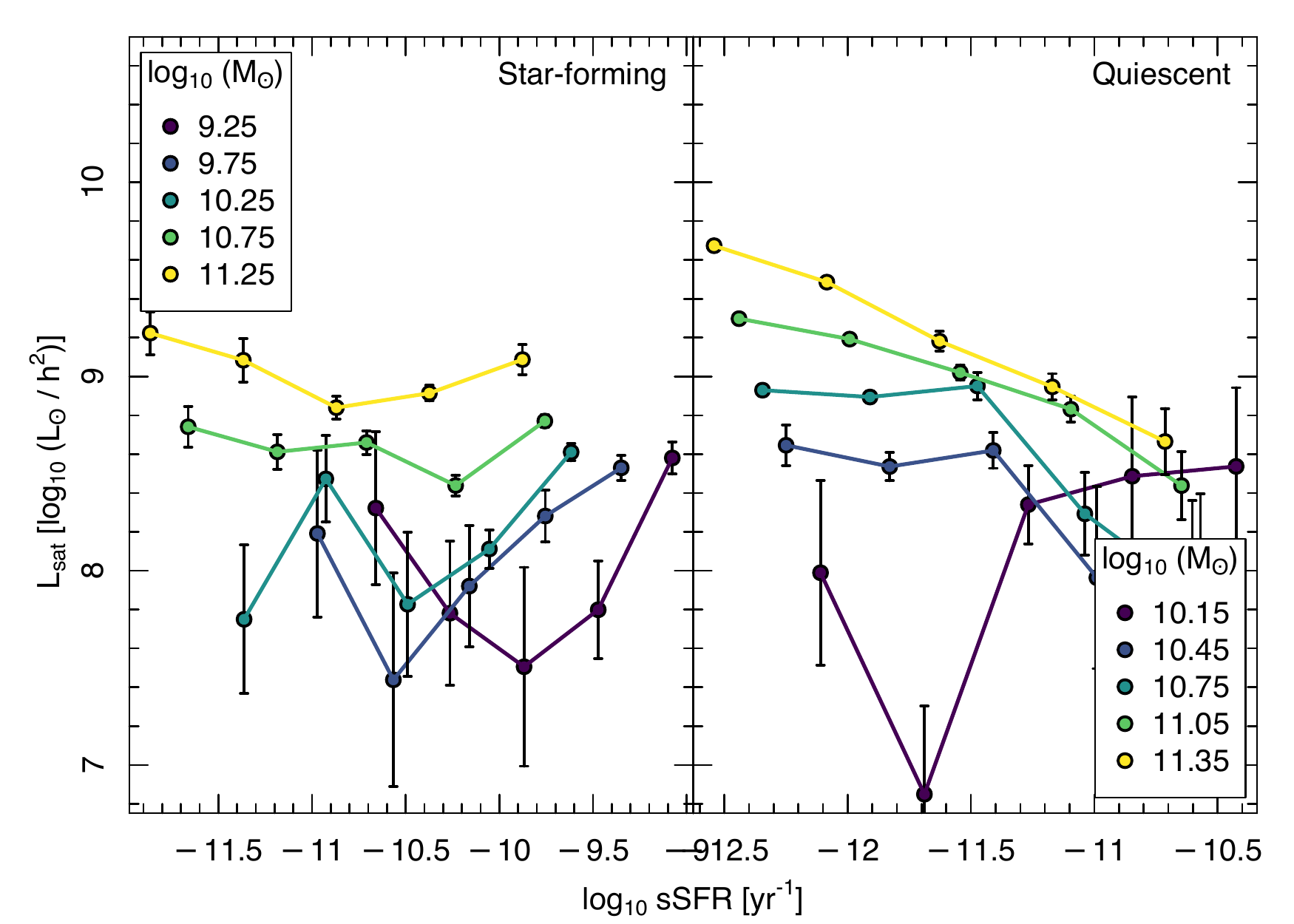}
	\includegraphics[width=0.495\textwidth]{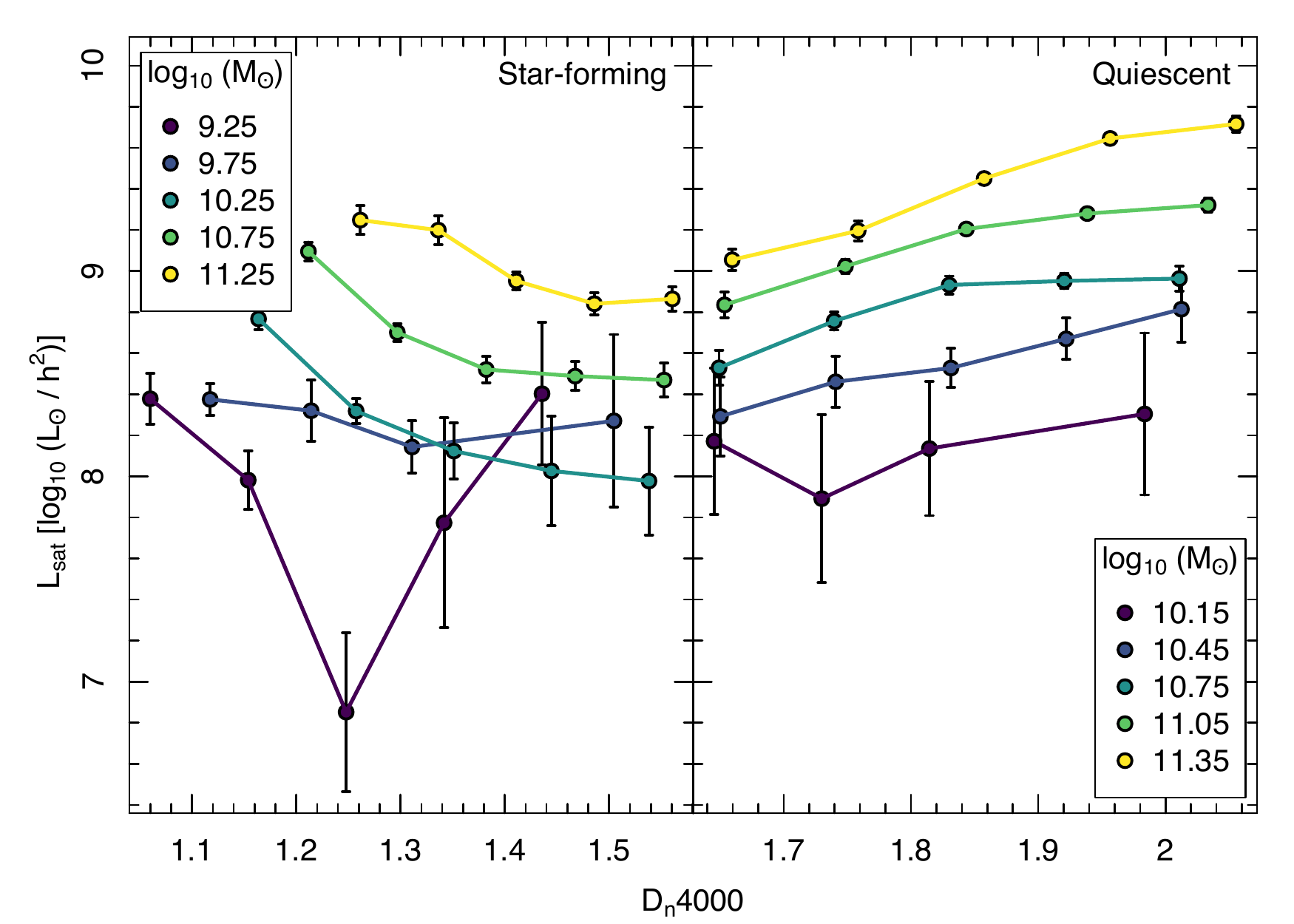}
	\includegraphics[width=0.495\textwidth]{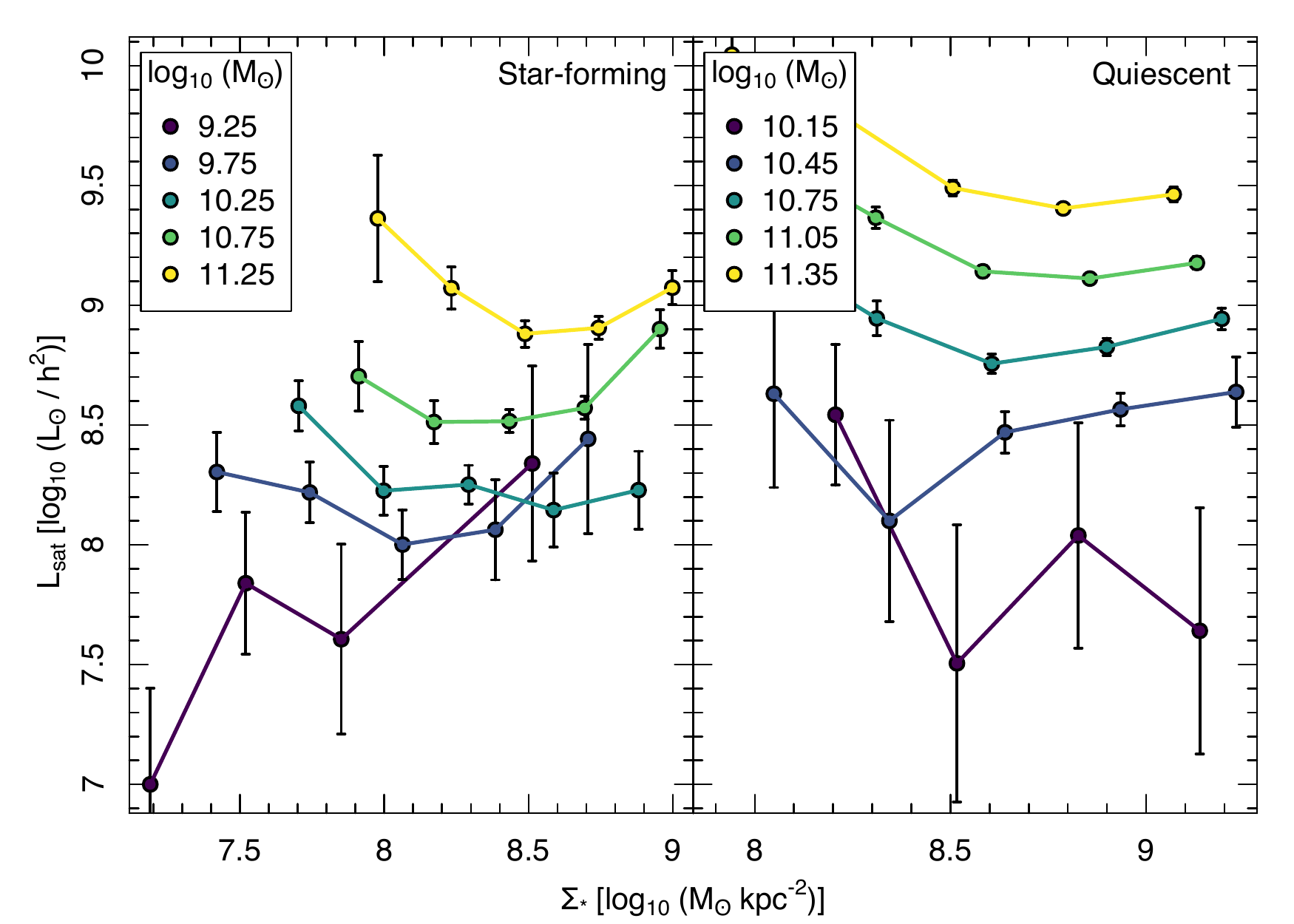}
	\includegraphics[width=0.495\textwidth]{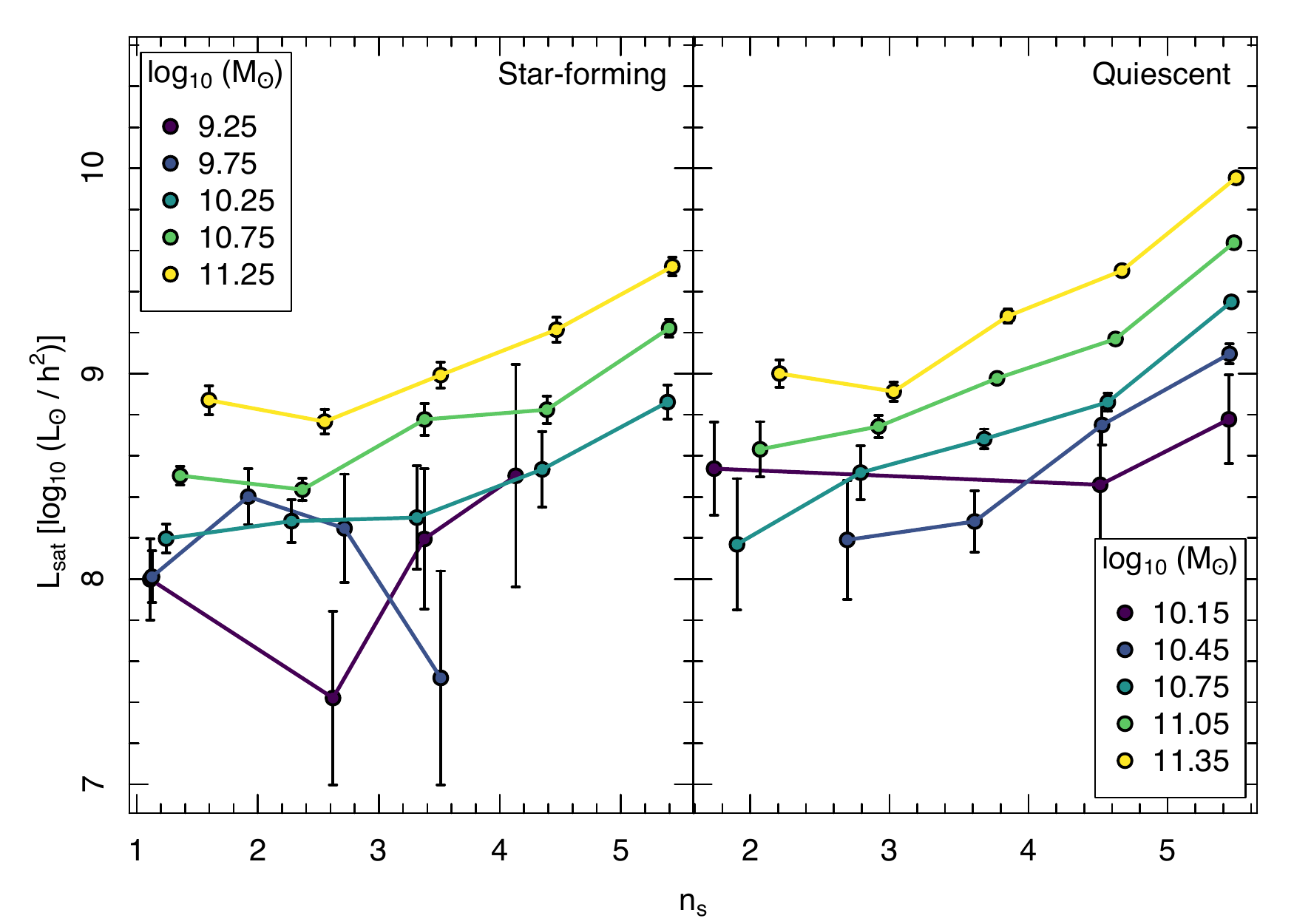}
	\caption{$\lsat$ plotted as a function of sSFR (top-left), D$_n4000$ (top-right), $\Sigma_*$ (bottom-left), and S\'ersic index (bottom-right) in bins of fixed stellar mass for populations of star-forming and quiescent galaxies as selected by their D$_n4000$ break strength. (left and right panels of each plot, respectively).}
	\label{fig:LsatRvB2}
\end{figure*}

Galaxies are known to differ in sizes and other properties within their different sub-populations. Red galaxies, for example, are systematically larger than blue galaxies. It is therefore important for us to check that the trends we have seen so far between galaxy properties and $\lsat$ are simply not being caused by the interrelationship between galaxy colour and other properties. In Figures \ref{fig:LsatRvB1} and \ref{fig:LsatRvB2} we display relationships between $\lsat$ and galaxy properties in bins of fixed stellar mass (i.e. the right-hand panels in the plots in Figures \ref{fig:LsatParms1} and \ref{fig:LsatParms2}) for galaxies with D$_n4000$ above and below 1.6. It is informative to observe the increase in correlation between different sub-populations; specifically in the case of $\sigma_v$ and $\reff$ where the mutual information with $\lsat$ is only higher than that of stellar mass in one sub-population at any given cut (e.g. blue galaxies and quiescent galaxies for $\sigma_v$ and red galaxies for $\reff$). It is also informative to see the different ranges in $\lsat$ that are occupied by these different sub-populations: for example, we can clearly see that red (quiescent) galaxies show higher values of $\lsat$ as a function of $M_r$ at fixed stellar mass compared to blue (star-forming) galaxies; suggesting that this sub-population is tracking large red and dead galaxies at the centers of halos that have accumulated large populations of satellites (this can also be seen clearly in the bottom right panel of Figure \ref{fig:LsatRvB1} where $\lsat$ is higher for galaxies with D$_n4000 \geq 1.6$).

\subsection{Mutual information}
\label{sec:MI}

While it is possible to see the correlation between $\lsat$ and the parameters we have discussed in the preceding section, we wish to objectively quantify the relative correlation strengths between different galaxy properties and $\lsat$. The determination of the amount of correlation between pairs (or more) of parameters is a widely studied problem, as are techniques by which data with large dimensionality can be reduced to a set of fundamental parameters (i.e. principal component analysis, or PCA). While PCA has been successfully applied to studies of galaxy evolution in the past (see e.g. \citealp{Woo2008}), this technique is best suited for `wide' data; i.e. data that consist of many observations for a relatively small number of objects. In this work we have `deep' data consisting of very many galaxies, but with a very noisy measurement of $\lsat$.

We choose instead to examine the mutual information of $\lsat$ with respect to stellar mass and the other parameters we have discussed. Mutual information (hereafter MI) is formally defined as the mutual dependence between two random variables $X$ and $Y$; more specifically, it quantifies the amount of information that can be gained on $Y$ given knowledge or observations of $X$ and is generally presented in units of bits. If $X$ and $Y$ are completely independent of each other (i.e. knowledge of $X$ in no way reduces the uncertainty associated with $Y$ or vice-versa), then the MI of $X$ and $Y$, written as $I(X;\,Y)$ is 0. In Appendix \ref{sec:MIapp} we discuss the theoretical underpinning of mutual information and provide examples of its calculation for simple data. 

\begin{table*}
	\centering
	\begin{tabular}{rcccc}
		\hline
	 	&Full sample&Quiescent&Star-forming&$\sigma_{\mathrm{jack}} \times 10^{-4}$\\ 
	 	\hline
		$M_*$ & 0.070 & 0.053 & 0.047 & $6.5$\\ 
		\hline
	  	$\reff$ & 0.052 & \textbf{0.058} & 0.041 & $6.6$\\ 
	  	$\sigma_v$ & 0.069 & 0.046 & \textbf{0.054} & $4.7$\\ 
	  	c$_{90/50}$ & 0.044 & 0.012 & 0.040 & $4.9$\\ 
	  	$M_r$ & \textbf{0.077} & \textbf{0.063} & \textbf{0.055} & 6.5\\ 
	  	D$_n 4000$ & 0.032 & 0.007 & 0.015 & 3.4\\ 
	  	sSFR & 0.037 & 0.012 & 0.021 & 3.4\\ 
	  	$\Sigma_*$ & 0.020 & 0.025 & 0.015 & 3.9\\ 
	  	$n_s$ & 0.050 & 0.012 & \textbf{0.049} & 7.1\\ 
	  	$12 + \log\,[\mathrm{O/H}]$ & 0.042 & 0.043 & 0.019 & 12.1\\ 
	   	\hline
	\end{tabular}
	\caption{Pairwise, bias-corrected mutual information values between $\lsat$ and the parameters listed in the left hand column for the whole central SDSS galaxy sample, as well for only for galaxies above and below D$_n 4000 = 1.6$. In this global view, only $r$-band absolute magnitude exceeds the correlation of stellar mass with $\lsat$ (as shown by the bold-faced values exceeding the MI values of stellar mass and $\lsat$). However, subdividing the galaxy population into sub-populations reveals cases where some parameters do perform better; e.g. $\reff$ for `red' galaxies; and $\sigma_v$ and S\'ersic index for `blue' galaxies. These cases have been printed in bold for clarity. Error estimates on these mutual information estimates are given in the final column are based on 10 jackknife resamples, and are of the order of $1\%$.}
	\label{tab:MIglobal}
\end{table*}

For the purposes of our work, we are interested in quantifying how much information there is to be gained on $\lsat$ by looking at galaxy parameters. In other words, we are specifically interested in finding out which galaxy properties of a central galaxy can best be used to predict its total satellite luminosity. In Table \ref{tab:MIglobal} we show the mutual information between $\lsat$ and other galaxy parameters across the entire sample of central SDSS galaxies, as well as subdivided by D$_n 4000$. When viewed globally, we see that only $M_r$ provides more information on $\lsat$ than stellar mass; however, we note that without performing any kind of binning, the global distribution of $\lsat$ is very noisy, which will necessarily reduce any amount of information shared between it and another parameter. When we subdivide the global population into two sub-categories: above and below D$_n 4000 = 1.6$, we do note that there are a number of cases where other parameters besides $M_r$ provide more information on $\lsat$ than stellar mass: $\reff$ for star-forming galaxies, and $\sigma_v$ and S\'ersic index for quiescent galaxies. In all of these cases the mutual information for $\lsat$ and these parameters exceeds that of $\lsat$ and stellar mass. We estimate the standard error on all mutual information estimates to be of the order of $1\%$ (i.e. $\sim 5 \times 10^{-5}$) based on recomputing all estimates for 10 jackknife resamples of the data. To quantify the influence of uncertainties in each galaxy property we have examined, we conduct a further 10 computations of mutual information where each galaxy property is perturbed by some amount drawn from a Gaussian distribution whose standard deviation is equal to the uncertainty in that parameter for that galaxy. In all cases, we find that the error estimates are too small to change which parameters perform better than stellar mass. In Appendix \ref{sec:MIapp} we interpret the significance of these mutual information values based on artificial data. Finally, we also include mutual information estimates for metallicities estimated from SDSS spectra taken from the MPA-JHU catalogue; however, these parameters are only given for approximately 14,000 galaxies in our sample. Due to this small sample size and its associated larger uncertainty, we only present this value in this table and opt not to investigate this parameter further in figures. The implication of this result is that when splitting by star-forming and quenched galaxies, $M_r$, $\reff$, $\sigma_v$, and $n_s$ carry more information on $\lsat$ than stellar mass; in other words, these parameters correlate more strongly with $\lsat$ than stellar mass.

\begin{figure*}
	\centering
	\includegraphics[width=1.0\textwidth]{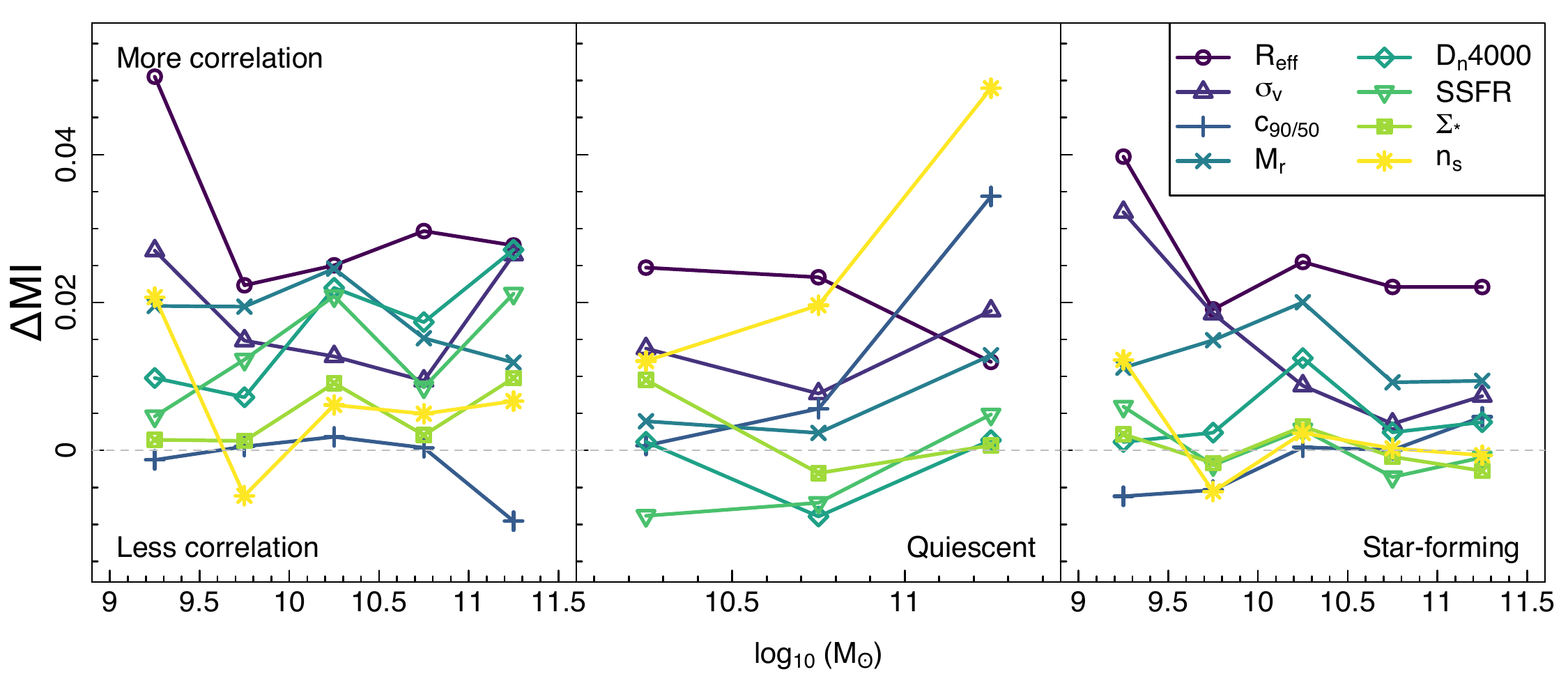}
	\caption{Change in mutual information between $\lsat$ and stellar mass and $\lsat$ and another parameter, shown as different symbols, in bins of stellar mass for the full sample of central galaxies (left); red galaxies (middle); and blue galaxies (right). See text for a definition of $\Delta \mathrm{MI}$ and the appendix for additional details of mutual information. Within this parameter space, a positive $\Delta \mathrm{MI}$ indicates that there is greater correlation between $\lsat$ and the parameter in question than there is between $\lsat$ and stellar mass. Most parameters show higher values of MI with $\lsat$ than stellar mass. Even for parameters like c$_{90/50}$ that do not perform as well, values of $\Delta \mathrm{MI}$ remain close to 0, suggesting that these parameters are on par with stellar mass within these mass bins. Note that we truncate masses below $M_* = 10^{10}$ M$_\odot$ for red galaxies due to low sample size below that mass.}
	\label{fig:deltaMI}
\end{figure*}

In Figures \ref{fig:LsatParms1} and \ref{fig:LsatParms2} we showed how $\lsat$ varies as a function of different parameters in bins of fixed stellar mass. We have already shown that the mutual information between $\lsat$ and these parameters provides some insight into which galaxy properties correlate most with $\lsat$, but it is also instructive to further examine how the mutual information varies as a function of stellar mass. Specifically, we wish to know how much more information we can gain on $\lsat$ by examining different parameters at fixed stellar mass. We quantify this by computing the following value:

\begin{equation}
\Delta \mathrm{MI} \equiv I(\lsat;\,X) - I(\lsat;\,M_{\odot})
\end{equation}

\noindent where $I(\lsat;\,M_{\odot})$ is the mutual information of $\lsat$ and stellar mass; and $I(\lsat;\,X)$ is the mutual information of $\lsat$ and another parameter, $X$. Both $I(\lsat;\,M_{\odot})$ and $I(\lsat;\,X)$ are bias-corrected via jackknife resampling. If $\Delta \mathrm{MI} > 0$, then knowledge of parameter $X$ provides more information on $\lsat$ than stellar mass. Figure \ref{fig:deltaMI} shows values of $\Delta \mathrm{MI}$ computed in bins of stellar mass for all of the parameters shown in the preceding section for the full sample of galaxies, as well as galaxies split by their D$_n4000$. We have set bin widths to be equal in size to the stellar mass bin widths of the figures shown in the preceding section. From this figure it is evident that across almost all stellar mass bins, stellar mass has a lower mutual information with $\lsat$ than a majority of secondary galaxy parameters. $\Delta \mathrm{MI}$ is greater than 0 for most parameters and performs best for $\reff$ and $\sigma_v$, which is consistent with what we have already shown. For other parameters such as c$_{90/50}$, $\Delta \mathrm{MI}$ is close to or below 0, indicating that that parameter performs equally as well as stellar mass as a predictor of $\lsat$. For parameters that perform better at higher stellar masses, Figure \ref{fig:deltaMI} shows results that are consistent: $\Delta$MI increases gradually to positive values at higher stellar masses. Across the entire stellar mass range, effective radius, r-band magnitude, and velocity dispersion perform best in this metric, with parameters like S\'ersic index and concentration performing poorest. We do not see significant differences in $\Delta \mathrm{MI}$ between the full sample and blue galaxies; but noter that for red galaxies we see a very strong signal for S\`{e}rsic index, c$_{90/50}$, and $\sigma_v$ for high mass red galaxies, likely due to the very homogeneous nature of this population at such high masses.

\section{Discussion}
\label{sec:discussion}

\begin{figure*}
	\centering
	\includegraphics[width=1\textwidth]{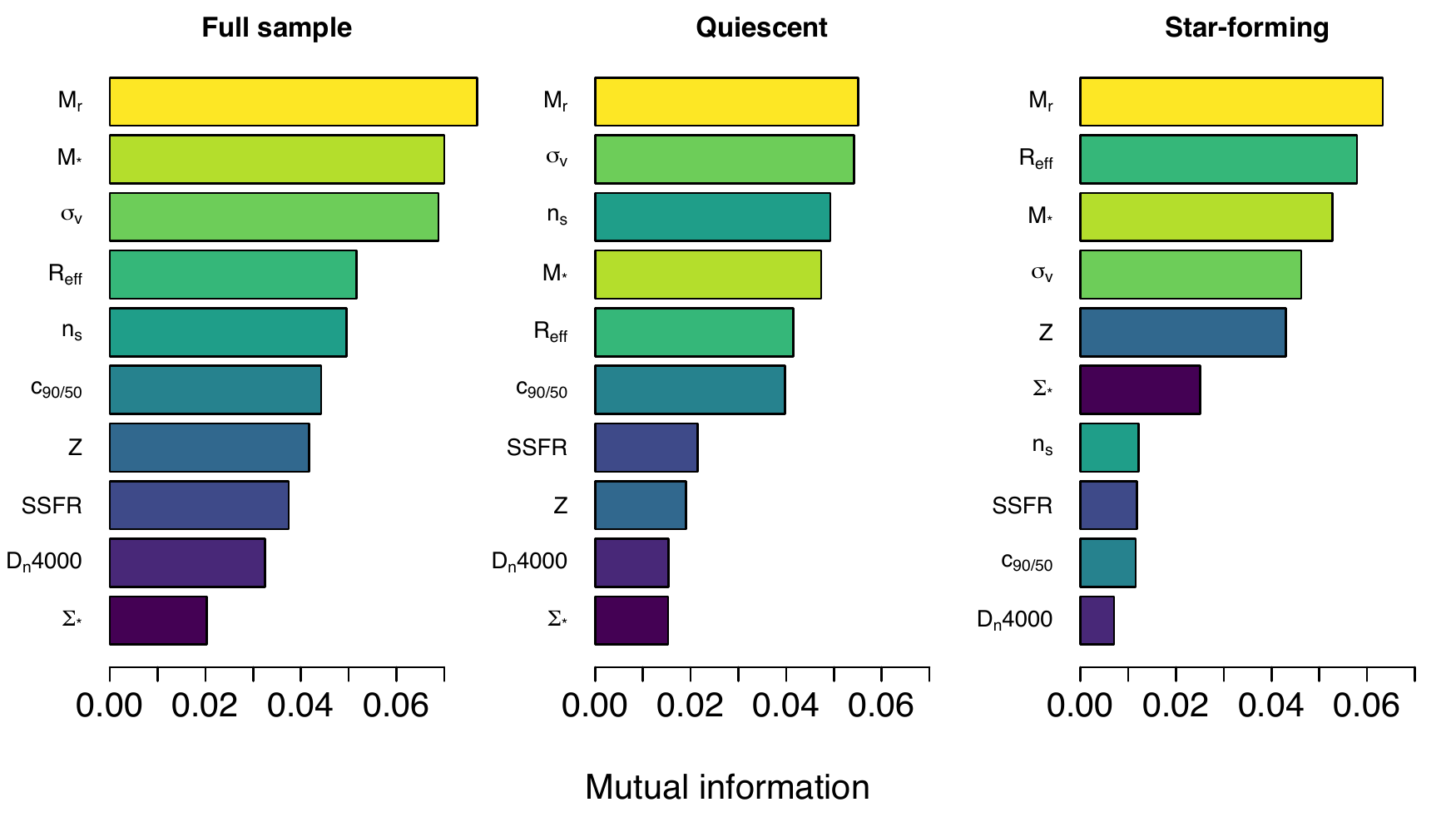}
	\caption{Barplots presenting a visual summary of the information presented in Table \ref{tab:MIglobal} of mutual information values between $\lsat$ and other galaxy properties discussed in this paper (the parameter Z denotes metallicity ($12 + \log_{10}\,$[O/H]). The left panel displays these values for the full sample, while the middle and right plots show MI values for star-forming and quiescent galaxies, as selected by their D$_n4000$ break strength. In each panel, parameters have been arranged in descending MI order, and each bar retains a consistent color across all panels for ease of viewing. The uncertainties for these MI values is given in \ref{tab:MIglobal}; we note that the uncertainties are too small to change the rank ordering of galaxy properties shown in this Figure.}
	\label{fig:MIsummary}
\end{figure*}

In the preceding section we have presented the relationship between $\lsat$ and numerous galaxy properties, as well as examined potential secondary biases that may be influencing these correlations. The mutual information between $\lsat$ and these parameters shows that when one considers the full galaxy sample without any binning at fixed stellar mass, $M_r$ contains the largest amount of information on $\lsat$. Figure \ref{fig:MIsummary} presents a visual summary of the data shown in Table \ref{tab:MIglobal}, in which we displayed the mutual information of $\lsat$ and the galaxy properties examined in this paper. Globally, $M_r$ always shows higher mutual information with $\lsat$ than stellar mass, but in quiescent and star-forming galaxies we see that $\reff$, $\sigma_v$ and $n_s$ contain more information as well. Knowing that $M_r$ and $M_*$ show the strongest correlation with $\lsat$, we can interpret the result of Figure \ref{fig:deltaMI} where we see a different rank ordering of the mutual information of galaxy properties as a function of stellar mass. Specifically, $\reff$ and $\sigma_v$ correlate more strongly than M$_r$ at fixed stellar mass, but M$_r$ has the highest mutual information with $\lsat$ for the full sample. This change in rank ordering is due to the fact that $M_r$ and $M_*$ have the strongest correlation with each other; so when we plot mutual information at fixed stellar mass we are seeing secondary correlations between these parameters and $\lsat$. It is notable that the mutual information of $n_s$ ranks low for the full galaxy population, but becomes one of the strongest signals for quiescent galaxies when binned by stellar mass; and one of the weaker ones for star-forming galaxies. We see a similar jump in rank for concentration, and note that there is likely a strong link between these two structural parameters.

In Figure \ref{fig:densgrid} we examine the relationship between each parameter and local density at fixed stellar mass. We find that in the case of quiescent galaxies, no observable parameter with the exception of sSFR shows strong trends with density. For star-forming galaxies, on the other hand, we see a strong correlation between $\reff$, $\sigma_v$, and $n_s$ with density. As discussed previously, the relationship between each observable parameter and density is key to determining if the relationship between that parameter and $\lsat$ is driven by halo mass or halo formation history; which in turn informs how much we can infer about the mass of a halo based on that observable property of the central galaxy. Based on the results in Figure \ref{fig:densgrid} for quiescent galaxies, we can infer  that the relationship between observables and $\lsat$ is driven by $M_h$ because we do not see density trends with these observables. This is not the case for star-forming galaxies, where parameters like $\reff$ and $\sigma_v$ show anti-correlations with density, meaning that halo formation history likely plays a role in their relationship with $\lsat$. The amplitude of the slope of the $\reff$-density relationship is consistent with that seen in a theoretical model where a parameter is explicitly tuned to anti-correlate with halo formation history (see Figure 4 in T19). The same is likely true for $\sigma_v$, though the amplitude of the $\sigma_v$-density slope is lower. Detailed modeling of galaxies and halos will confirm to what extent halo formation history or halo mass play a role in shaping the trend between these parameters and $\lsat$; we leave this to a future work.

One exception to the aforementioned degenaracy for star-forming galaxies with observables that correlate with density is the way $n_s$ increases as a function of density for star-forming galaxies in Figure \ref{fig:densgrid}. These results are qualitatively consistent with the marked correlation function analysis of \citet{Calderon2018}, who also found correlations between $n_s$ and large-scale density. However, $n_s$ correlates positively with $\lsat$ at fixed stellar mass in Figure \ref{fig:LsatRvB2}. If the correlation between $\lsat$ and $n_s$ is driven by halo formation history, then one would expect to find galaxies with high $n_s$ at lower densities; since older halos are found in less dense environments. That we see an inverse trend with density and $n_s$ in Figure \ref{fig:densgrid} suggests that the relationship between $\lsat$ and $n_s$ may in fact be driven by halo mass, rather than by halo formation history. One argument in favor of the positive correlation between $n_s$ and density being driven by $M_h$ is that the scatter of $M_h$ at fixed $M_*$ becomes very large at high stellar mass \citep{Wechsler2018}, and this may be being reflected in the results in Figure \ref{fig:densgrid}, where the trend between $n_s$ and density increases as a function of stellar mass.

T19 show that the single biggest systematic in $\lsat$ measurements is the accurate selection of central galaxies; which is a point we revisit when considering the results seen in Figure \ref{fig:lsatdens}. As discussed in the body of the paper, we would have seen a positive correlation between $\lsat$ and density if our sample had significant contamination from misclassified central galaxies; however, we observe a downward trend between $\lsat$ and density. Based on these considerations it is unlikely that the relationships we see between $\lsat$ and the various parameters we have examined in this work are driven by systematics unrelated to the connection between central galaxy properties and their satellite population. In other words, regardless of whether or not the relationship between these secondary galaxy properties and $\lsat$ is driven by halo mass or halo formation history, $\lsat$ is a clear observable probe of the properties of the host halo of a galaxy.

Finally, we note that all of our results are sensitive to the choice of stellar masses that we use in our analysis; and there are numerous stellar mass catalogues available for the galaxy sample we have used. In a future paper we will examine the relationship between $\lsat$ and publicly available stellar mass catalogues for the SDSS to test how different stellar mass estiamtes correlate with dark matter halo properties.

\section{Conclusion and summary}
\label{sec:conclusion}

We have utilized spectroscopic data from SDSS and photometric imaging data from the LS to examine the relationship between the properties of 117,966 central galaxies within $ z= 0.15$ and the total luminosity of their satellites. The total satellite luminosity $\lsat$ of a galaxy has been shown to scale linearly with its stellar mass (T19), and in this work we have shown that in many cases, particularly when galaxies are binned by their stellar masses, secondary properties of galaxies such as their size and luminosity correlate even more strongly with $\lsat$ than their stellar masses. Local density can be a potential source of bias. We devise a density metric that is derived from number counts of neighbours within 10 h$^{-1}$ Mpc around each central galaxy (corrected for edge effects), and is also shown to be unbiased in bins of fixed redshift and stellar mass. In Figures \ref{fig:lsatdens} and \ref{fig:densgrid} we show that there are no significant relationships between density and $\lsat$ or the other parameters we examine.  

Most notably, despite the overall noise in $\lsat$ on an individual galaxy basis, $r$-band absolute magnitude consistently carries more information on $\lsat$ than stellar mass. While this is evident from viewing Figures \ref{fig:LsatParms1} and \ref{fig:LsatParms2}, we explicitly quantify it by computing the mutual information of $\lsat$ and stellar mass and compare it to the mutual information of $\lsat$ and all other parameters. These results, summarized in Table \ref{tab:MIglobal} and Figure \ref{fig:deltaMI} for mutual information values computed globally and at fixed stellar mass, show that with a few exceptions, secondary parameters provide important constraints on $\lsat$. We extend our analysis to examine how $\lsat$ varies when binned along two parameters, and show that not all parameters exact an equal influence on $\lsat$. A hierarchy of parameters' correlation to $\lsat$ is further established by quantifying the mutual information shared between $\lsat$ and each parameter. Discussed further in Appendix \ref{sec:MI}, the mutual information of two parameters measures the amount of information gained on the second parameter with knowledge of the first. We compute the mutual information between $\lsat$ and all other galaxy properties in our study and show their results in Table \ref{tab:MIglobal} and Figures \ref{fig:MIsummary} and \ref{fig:deltaMI}. These results show that in all cases, the absolute r-band magnitude of a galaxy M$_r$ carries more information on $\lsat$ than its stellar mass, which one might expect to be the property that informs total satellite luminosity the most. For red, quiescent galaxies with D$_n4000\geq 1.6$ we see that $\sigma_v$ and $n_s$ also carry more information on $\lsat$; while for star-forming blue galaxies $\reff$ also has more mutual information with $\lsat$ than $M_*$. The interpretation of these results is that the correlation between $\lsat$ and M$_r$ is strongest, and that knowledge of the latter is the best way to predict the former quantity.

As discussed in T19, the relationship between $\lsat$ and a secondary galaxy parameter can be driven both by the mass of the host halo $M_h$, as well as its formation history. One way to disambiguate between these two scenarios is to examine the relationship between a given secondary galaxy parameter and the large-scale environment of that galaxy. T19 show that in cases where a secondary galaxy parameter does not correlate with large-scale density, the relationship between that secondary parameter and $\lsat$ is driven entirely by $M_h$. In Figure \ref{fig:densgrid} we show that in quiescent galaxies, no secondary galaxy properties show any dependence on $\delta_\sigma$, meaning that the relationship between these properties and $\lsat$ for quiescent galaxies is driven entirely by $M_h$. Things are less clear for star-forming galaxies, where $\reff$, $\sigma_v$, and S\'ersic index all show a correlation with $\delta_\sigma$, implying that halo formation history may influence the relationship between these properties and $\lsat$. The relative influence of M$_h$ and halo formation history, as well as quantification of any observational systematic biases in our central galaxy selection requires detailed modeling that we leave to a future paper. We do note, however, that if the link between $\lsat$ and galaxy properties for star-forming galaxies is influenced by formation history rather than halo mass, it is required that the quenching process remove this correlation.

Our observation that $\lsat$ correlates strongly with $\reff$ can be interpreted within the context of previous works which examine the degree to which the relationship between the size of a galaxy and its host halo mass is influenced by secondary parameters. Recent work by \citet{Desmond2017}, \citet{Hearin2017}, and \citet{Somerville2018} using various techniques has shown that there is an anticorrelation between galaxy size and halo size at fixed stellar mass, which is in contradiction with our results which show a positive correlation with $\lsat$ and $\reff$ at fixed stellar mass (particularly for the case of quiescent galaxies whose $\reff$ is shown not to correlate with $\delta_\sigma$ in Figure \ref{fig:densgrid}). On the other hand, work by \citet{Kravtsov2013} has established a linear relationship between galaxy half-mass radius and halo radius at fixed stellar mass for both star-forming and quiescent galaxies; more recent work by \citet{Huang2017} using an abundance matching framework at observational results from the CANDELS survey \citep{Galametz2013} echoes these results, finding a quasi-linear relationship between galaxy size and halo size for star-forming and quiescent galaxies. Both of these findings are in quantitatively line with our results.

$\lsat$ is a promising and important probe in quantifying the galaxy-halo connection, as it opens the door to improving our understanding of halo properties of galaxy groups down to lower mass thresholds than existing methods. The findings presented in this work extend the utility of this method by drawing a connection between numerous secondary galaxy parameters, some of which like $\sigma_v$ and $M_r$ can be directly measured from data without significant assumptions and processing, and $\lsat$. We hope to draw upon these correlations and incorporate them into groupfinding techniques in future works.

\section*{Acknowledgements}
MA wishes to thank Johann Brehmer for extremely useful discussions regarding mutual information estimators.

All analysis in figures in this manuscript have been prepared using the R programming language \citep{R-project}, making use of the following packages and libraries: \texttt{wesanderson} \citep{wesanderson}, \texttt{Cairo} \citep{Cairo}, \texttt{astro} \citep{astro}, \texttt{magicaxis} \citep{magicaxis}, \texttt{MASS} \citep{MASS}, \texttt{celestial} \citep{celestial}, \texttt{hyper.fit} \citep{hyper.fit}, \texttt{data.table} \citep{data.table}, \texttt{IDPmisc} \citep{IDPmisc}, \texttt{RColorBrewer} \citep{RColorBrewer}, \texttt{boot} \citep{bootpackage,bootbook}, \texttt{viridis} \citep{viridis}, \texttt{foreach} \citep{foreach}, \texttt{fields} \citep{fields}, and \texttt{MPMI} \citep{mpmi}.  

Funding for the SDSS and SDSS-II has been provided by the Alfred P. Sloan Foundation, the Participating Institutions, the National Science Foundation, the U.S. Department of Energy, the National Aeronautics and Space Administration, the Japanese Monbukagakusho, the Max Planck Society, and the Higher Education Funding Council for England. The SDSS Web Site is http://www.sdss.org/. 

The SDSS is managed by the Astrophysical Research Consortium for the Participating Institutions. The participating institutions are the American Museum of Natural History, Astrophysical Institute Potsdam, University of Basel, University of Cambridge, Case Western Reserve University, University of Chicago, Drexel University, Fermilab, the Institute for Advanced Study, the Japan Participation Group, Johns Hopkins University, the Joint Institute for Nuclear Astrophysics, the Kavli Institute for Particle Astrophysics and Cosmology, the Korean Scientist Group, the Chinese Academy of Sciences (LAMOST), Los Alamos National Laboratory, the Max-PlanckInstitute for Astronomy (MPIA), the Max-Planck-Institute for Astrophysics (MPA), New Mexico State University, Ohio State University, University of Pittsburgh, University of Portsmouth, Princeton University, the United States Naval Observatory, and the University of Washington.

The DESI Legacy Imaging Surveys consist of three individual and complementary projects: the Dark Energy Camera Legacy Survey (DECaLS; NOAO Proposal ID \# 2014B-0404; PIs: David Schlegel and Arjun Dey), the Beijing-Arizona Sky Survey (BASS; NOAO Proposal ID \# 2015A-0801; PIs: Zhou Xu and Xiaohui Fan), and the Mayall z-band Legacy Survey (MzLS; NOAO Proposal ID \# 2016A-0453; PI: Arjun Dey). DECaLS, BASS and MzLS together include data obtained, respectively, at the Blanco telescope, Cerro Tololo InterAmerican Observatory, National Optical Astronomy Observatory (NOAO); the Bok telescope, Steward Observatory, University of Arizona; and the Mayall telescope, Kitt Peak National Observatory, NOAO. The DESI Legacy Imaging Surveys project is honored to be permitted to conduct astronomical research on Iolkam Du’ag (Kitt Peak), a mountain with particular significance to the Tohono O’odham Nation.

NOAO is operated by the Association of Universities for Research in Astronomy (AURA) under a cooperative agreement with the National Science Foundation.

This project used data obtained with the Dark Energy Camera (DECam), which was constructed by the Dark Energy Survey (DES) collaboration. Funding for the DES Projects has been provided by the U.S. Department of Energy, the U.S. National Science Foundation, the Ministry of Science and Education of Spain, the Science and Technology Facilities Council of the United Kingdom, the Higher Education Funding Council for England, the National Center for Supercomputing Applications at the University of Illinois at Urbana-Champaign, the Kavli Institute of Cosmological Physics at the University of Chicago, Center for Cosmology and Astro-Particle Physics at the Ohio State University, the Mitchell Institute for Fundamental Physics and Astronomy at Texas A\&M University, Financiadora de Estudos e Projetos, Fundacao Carlos Chagas Filho de Amparo, Financiadora de Estudos e Projetos, Fundacao Carlos Chagas Filho de Amparo a Pesquisa do Estado do Rio de Janeiro, Conselho Nacional de Desenvolvimento Cientifico e Tecnologico and the Ministerio da Ciencia, Tecnologia e Inovacao, the Deutsche Forschungsgemeinschaft and the Collaborating Institutions in the Dark Energy Survey. The Collaborating Institutions are Argonne National Laboratory, the University of California at Santa Cruz, the University of Cambridge, Centro de Investigaciones Energeticas, Medioambientales y Tecnologicas-Madrid, the University of Chicago, University College London, the DES-Brazil Consortium, the University of Edinburgh, the Eidgenossische Technische Hochschule (ETH) Zurich, Fermi National Accelerator Laboratory, the University of Illinois at Urbana-Champaign, the Institut de Ciencies de l’Espai (IEEC/CSIC), the Institut de Fisica d’Altes Energies, Lawrence Berkeley National Laboratory, the Ludwig-Maximilians Universitat Munchen and the associated Excellence Cluster Universe, the University of Michigan, the National Optical Astronomy Observatory, the University of Nottingham, the Ohio State University, the University of Pennsylvania, the University of Portsmouth, SLAC National Accelerator Laboratory, Stanford University, the University of Sussex, and Texas A\&M University.

BASS is a key project of the Telescope Access Program (TAP), which has been funded by the National Astronomical Observatories of China, the Chinese Academy of Sciences (the Strategic Priority Research Program ``The Emergence of Cosmological Structures'' Grant \# XDB09000000), and the Special Fund for Astronomy from the Ministry of Finance. The BASS is also supported by the External Cooperation Program of Chinese Academy of Sciences (Grant \# 114A11KYSB20160057), and Chinese National Natural Science Foundation (Grant \# 11433005).

The DESI Legacy Imaging Surveys team makes use of data products from the Near-Earth Object Wide-field Infrared Survey Explorer (NEOWISE), which is a project of the Jet Propulsion Laboratory/California Institute of Technology. NEOWISE is funded by the National Aeronautics and Space Administration.

The DESI Legacy Imaging Surveys imaging of the DESI footprint is supported by the Director, Office of Science, Office of High Energy Physics of the U.S. Department of Energy under Contract No. DE-AC02-05CH1123, by the National Energy Research Scientific Computing Center, a DOE Office of Science User Facility under the same contract; and by the U.S. National Science Foundation, Division of Astronomical Sciences under Contract No. AST-0950945 to NOAO.

%%%%%%%%%%%%%%%%%%%%%%%%%%%%%%%%%%%%%%%%%%%%%%%%%%

%%%%%%%%%%%%%%%%%%%% REFERENCES %%%%%%%%%%%%%%%%%%

% The best way to enter references is to use BibTeX:

\bibliographystyle{mnras}
\bibliography{lsat,R-Code}

%%%%%%%%%%%%%%%%%%%%%%%%%%%%%%%%%%%%%%%%%%%%%%%%%%

%%%%%%%%%%%%%%%%% APPENDICES %%%%%%%%%%%%%%%%%%%%%

\appendix

\section{The density metric $\delta_\sigma$}
\label{sec:densApp}

Avoiding systematic biases in galaxy density computations is crucial. In this Appendix we describe in further detail our reasoning and methodology behind computing $\delta_\sigma$. We begin by doing a simple count of galaxies in 10 Mpc spheres centered on each of our central SDSS galaxies. The background galaxy population that we count is the full NYU-VAGC spectroscopic galaxy catalogue. As mentioned in the main body of the text, we opt not to use a volume limited density defining population (as done in e.g. \citealp{Baldry2006,Brough2013}) because this drastically reduces the total number of background galaxies that we can count from, which in turn leads to much higher errors in $\rho$. 

\begin{figure}
	\centering
	\includegraphics[width=0.5\textwidth]{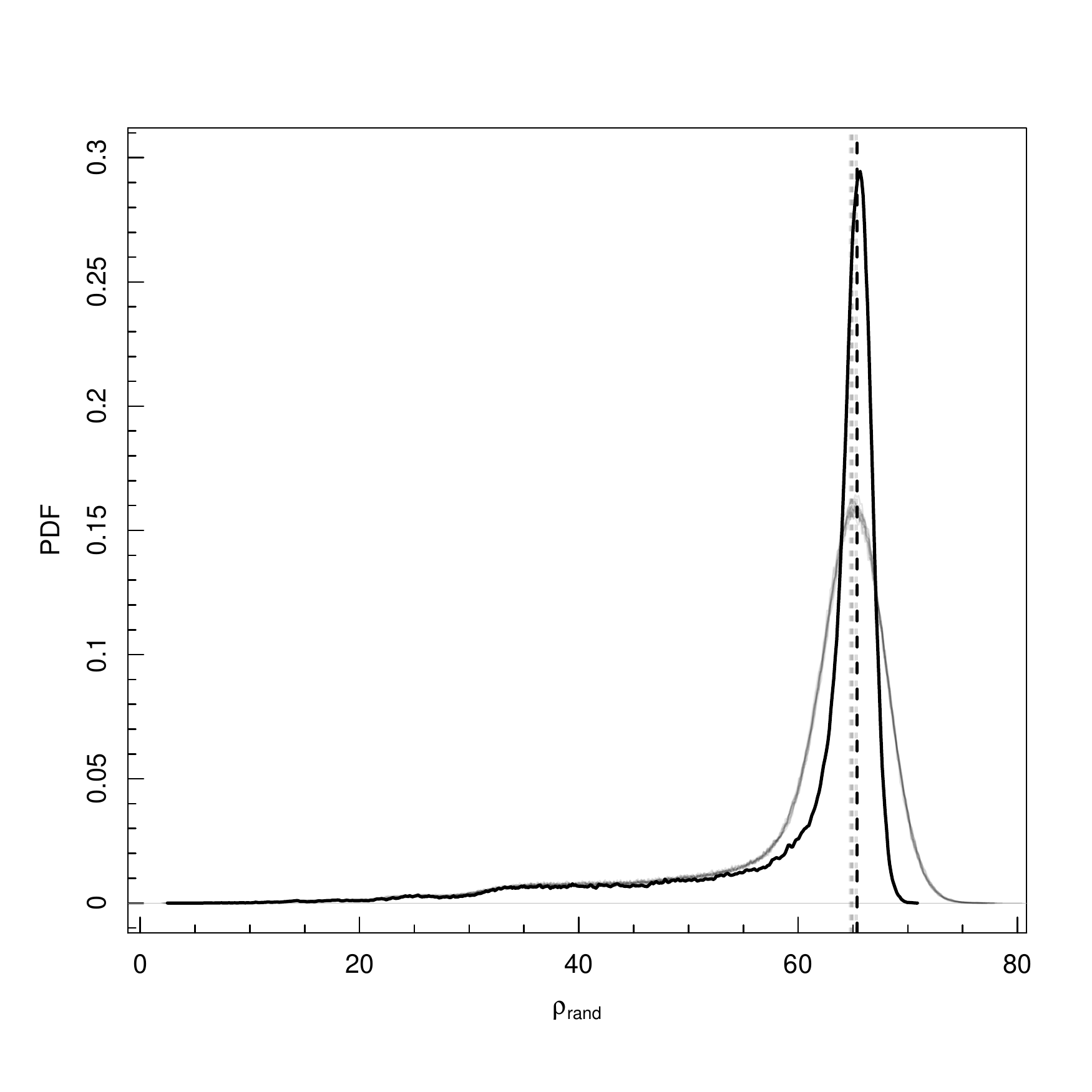}
	\caption{Probability density functions of the number density $\rho_{\mathrm{rand}}$ of random points within 10 Mpc of each central SDSS galaxy for all 10 random sub-catalogues (shown in grey) and across all $10^8$ randoms (shown in black). Vertical lines are placed at the mode of each distribution (65.38 for the full population).}
	\label{fig:randvDens}
\end{figure}

\begin{figure}
	\centering
	\includegraphics[width=0.5\textwidth]{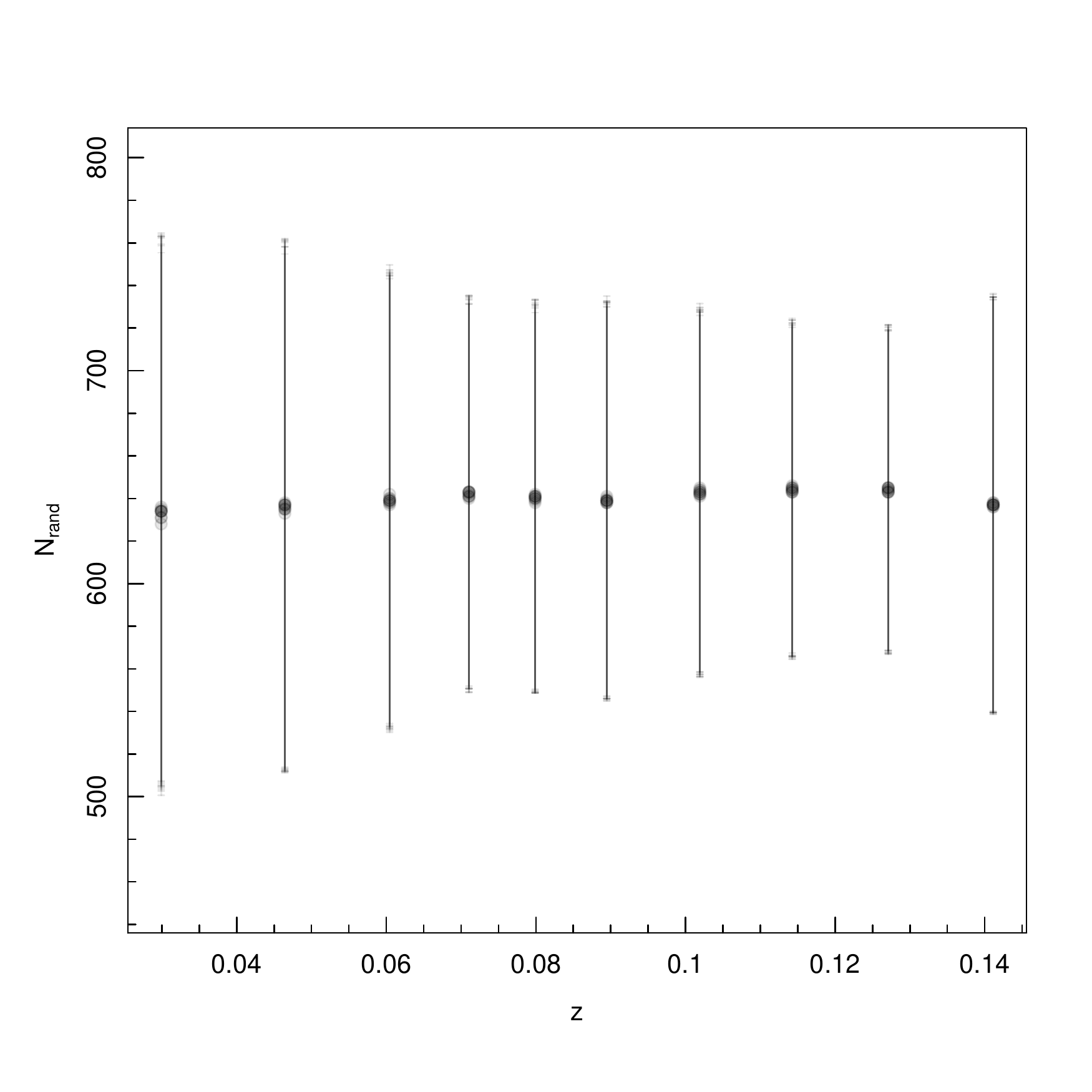}
	\caption{The number of random points $N_{\mathrm{rand}}$ within 10 Mpc of each central SDSS galaxy for all 10 random sub-catalogues (shown in grey) and across all $10^8$ randoms (shown in black), plotted as a function of redshift. Points show the median value of $N_{\mathrm{rand}}$ in each redshift bin, with error bars indicating the standard error on the median. We observe no change in $N_{\mathrm{rand}}$ as a function of redshfit, indicating that the random points have a uniform radial distribution.}
	\label{fig:randvZ}
\end{figure}

Regardless of how which sample of background galaxies is used for counting, all density measurements made with galaxy survey data will suffer from edge effects, with number counts artificially dropping near the survey's edge. For a geometrically simple survey this can be taken into account easily by defining the edges of the survey (either by ignoring galaxies within a certain distance of this edge, or weighting their number counts accordingly); however, in the case of our central SDSS galaxies, we are faced with a much more complex survey mask (see Figure \ref{fig:skymap}). Rather than define this complicated geometry, we take a set of $10^7$ points distributed randomly in longitude and latitude that match this geometry (these are obtained from the 6th and 7th data releases of the DESI Legacy Imaging Surveys) and spherically point it radially from $z = 0$ to $z = 0,.15$ in order to match our sample. For each random point we generate a distance $D$ such that $D = R_S U^{1/3}$ where $R_S$ is the maximum radius of our sphere (in this case 623.65 $h^{-1}$ Mpc, the comoving distance at $z = 0.15$ in our chosen cosmology) and $U$ is a uniform random number between 0 and 1. This algorithm distributes the random points radially such that there is a uniform distribution of points as a function of radius. By generating 10 such catalogues (assigning 10 random distances to each random point) we generate $10^8$ random points and count how many there are within 10 Mpc of every central SDSS galaxy in our sample. We show PDFs of the number density of $N_{\mathrm{rand}}$, $\rho_{\mathrm{rand}}$, in Figure \ref{fig:randvDens} where it can be seen that $N_{\mathrm{rand}}$ is uniform across all 10 random sub-catalogues; and Figure \ref{fig:randvZ} shows that there is no evolution in $N_{\mathrm{rand}}$ as a function of redshift. Knowing the typical number of random points around each central makes it trivial to identify which centrals are close to the survey edge. We therefore compute $\rho$ such that it is the number count of background NYU-VAGC galaxies around each central galaxy, weighted by $N_{\mathrm{rand}} / 65.38$ for that galaxy. 

\begin{figure*}
	\centering
	\includegraphics[width=1.0\textwidth]{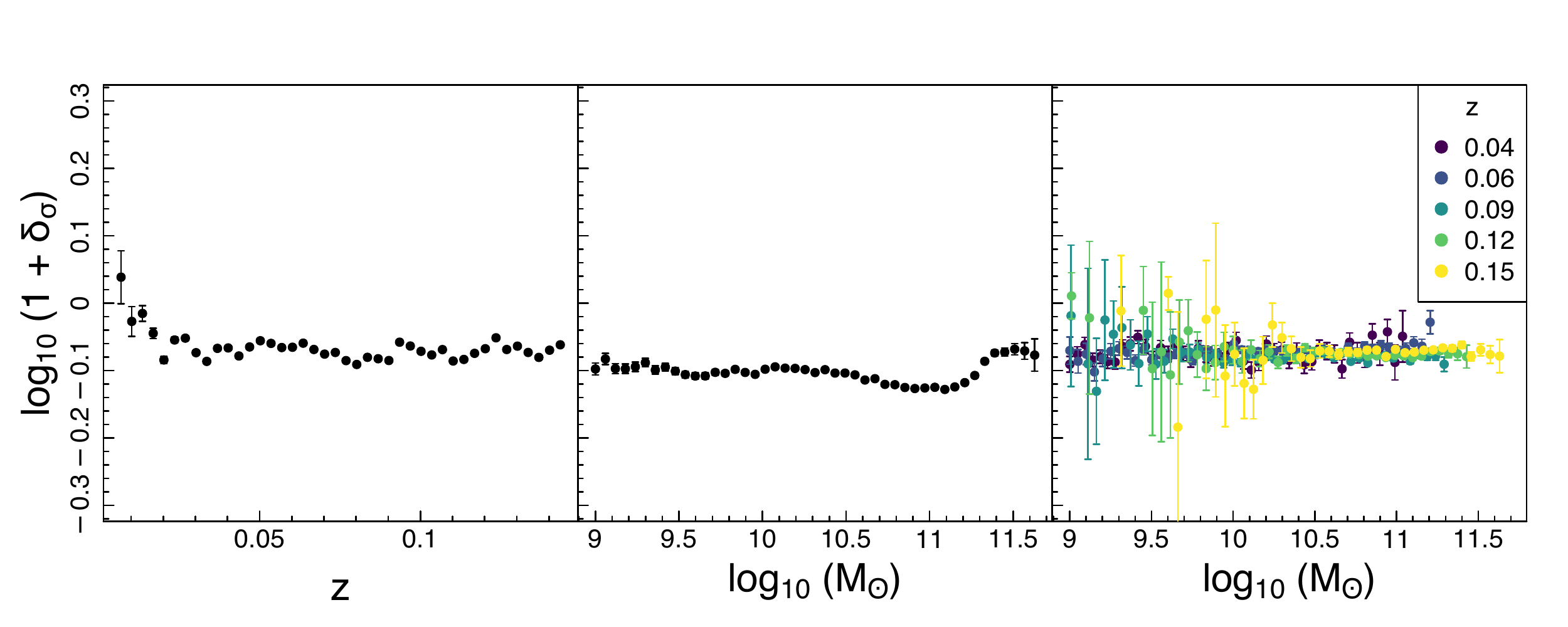}
	\caption{\emph{Left:} $1 + \log_{10}(1+\delta_\sigma)$ plotted as in bins of fixed redshift. \emph{Middle:} $1 + \log_{10}(1+\delta_\sigma)$ plotted in bins of fixed of stellar mass. \emph{Right:} $1 + \log_{10}(1+\delta_\sigma)$ plotted in bins of both fixed stellar mass and fixed redshift (as shown as the different colored lines). There is no visible bias in $1 + \log_{10}(1+\delta_\sigma)$ at fixed stellar mass and redshift. Each point shows the median value of $1 + \log_{10}(1+\delta_\sigma)$ in the bin, and the error bars show the standard error about the median.}
	\label{fig:checkRhoPlots}
\end{figure*}

With density estimates $\rho$ for each central galaxy weighted by their proximity to the edge, we can proceed to compute $\delta_\sigma$. In Figure \ref{fig:checkRhoPlots} we plot $1 + \log_{10}(1+\delta_\sigma)$ for galaxies binned by redshift, stellar mass, and both (left, central, and right panels respectively). From the right-most panel of this figure it is apparent that $1 + \log_{10}(1+\delta_\sigma)$ is an unbiased density estimator at fixed redshift and stellar mass, making it adequate for our purposes in studying how $\lsat$ varies as a function of density in various bins of redshift and stellar mass.

\section{Mutual information}
\label{sec:MIapp}

Formally, the mutual information between two random variables $X$ and $Y$ is defined as 

\begin{equation}
I(X;\,Y) = D_{\mathrm{KL}} (P_{(X,\,Y)}) \| P_X \otimes P_Y) 
\label{eqn:defMI}
\end{equation}

\noindent where $D_{KL}$ is a metric known as the Kullback-Leibler divergence (also referred to as the mutual entropy) of $X$ and $Y$. Given two distributions of continuous random variables $P$ and $Q$, the Kullback-Leibler divergence measures the difference between these two distributions in the following way:

\begin{equation}
D_{\mathrm{KL}}(P\|Q) = \int_{-\infty}^{\infty} p(x) \log \left( \frac{p(x)}{q(x)} \right) \ud x
\label{eqn:DKL}
\end{equation}

\noindent where $p(x)$ and $q(x)$ are the probability density functions of $P$ and $Q$. Recasting Eqn. \ref{eqn:defMI} with the definition for the Kullback-Leibler divergence given in Eqn. \ref{eqn:DKL} yields the following explicit definition of mutual information for a pair of continuous random variables $X$ and $Y$:

\begin{equation}
I(X;\,Y) = \int_y \int_x p_{(X,\,Y)}(x,\,y) \log \left( \frac{p_{(X,\,Y)}(x,\,y)}{p_X(x)p_Y(y)}\right) \ud x \ud y
\label{eqn:fullMI}
\end{equation}

\noindent where $p_{(X,\,Y)}$ is the joint probability distribution of $X$ and $Y$ and $p_X$ and $p_Y$ are the marginal probability distributions of $X$ and $Y$ respectively. Note that we use log base 2 in our computation, meaning that the units of I(X;\,Y) are bits. Considering the case where $X$ and $Y$ are independent, then $p_{(X,\,Y)}(x,\,y) = p_X(x) \times p_Y(y)$, which leads to a mutual information of 0 (as $\log 1 = 0$). Note that mutual information is symmetric; i.e. $I(X;\,Y) = I(Y;\,X)$. In this work, we compute mutual information using the algorithm of \citet{mpmi}, provided via the R package \texttt{MPMI}.

\begin{figure*}
	\centering
	\includegraphics[width=1.0\textwidth]{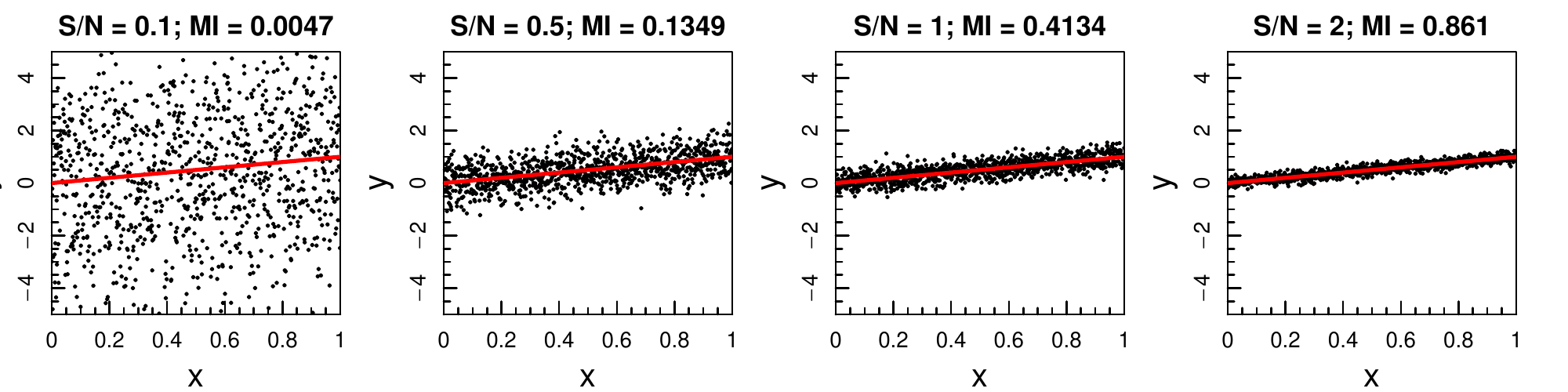}
	\caption{Mutual information estimates between two variables $x$ and $y$, with noise added to $y$ in differing amounts as shown by the signal-to-noise ratio given in the heading of each panel, next to the computed mutual information for the data shown. Points in $y$ are generated using Gaussian noise. In all panels, the red line drawn over the points indicates the 1-1 relation about which the points are generated.}
	\label{fig:MItest}
\end{figure*}

\begin{figure}
	\centering
	\includegraphics[width=0.5\textwidth]{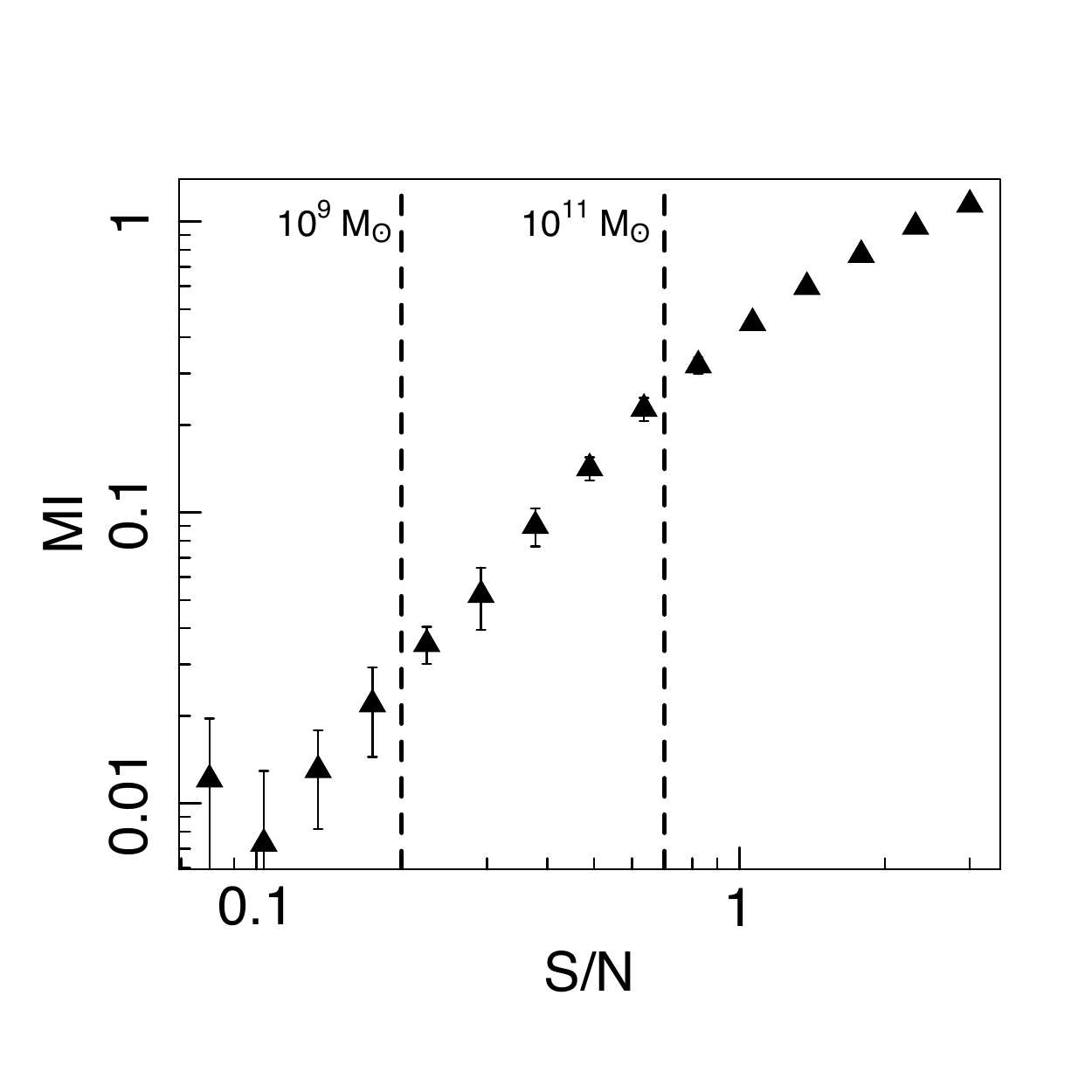}
	\caption{Mutual information as a function of S/N ratio as computed in Figure \ref{fig:MItest} for S/N ratios on $\lsat$ as a function of mass as reported in T19. The dashed vertical lines correpond to S/N ratios of $\lsat$ reported in T19 for galaxies with $M_* = 10^9$ and $10^{11}$ respectively from left to right. The points shown in this Figure are means for 20 realizations of random numbers, with standard errors shown.}
	\label{fig:MItest2}
\end{figure}

In order to better interpret the context and significance of mutual information estimates computed in this work, we compute mutual information estimates between two sets of generated data $x$ and $y$. These data are generated by adding Gaussian noise to $y$ such that we generate data with differing amounts of signal-to-noise (0.1, 0.5, 1, and 2) and are shown in Figure \ref{fig:MItest}. We add noise to $y$ to mimic the uncertainties in $\lsat$, which we consistently plot as the ordinate parameter when comparing to other galaxy properties. In each case, we compute the mutual information between $x$ and $y$ in exactly the same way as we do in Section \ref{sec:MI}; these values are shown on the title of each panel in Figure \ref{fig:MItest}. Predictably, MI increases as a function of S/N ratio. In Table \ref{tab:MIglobal} we report values for mutual information between 0.012 and 0.077; these values are consistent with the lower S/N mutual information values we compute in Figure \ref{fig:MItest} but are not as low as what we compute for a S/N ratio of 0.1. Generally, S/N ratios of $\sim 0.4$ produce MI estimates that are of the order of 0.05 to 0.08. Such ratios are entirely consistent with the inherent S/N ratio associated with $\lsat$ as reported in T19, who estimate inherent S/N ratios for $\lsat$ from simulations based on the total satellite luminosity of known group satellites versus the noise contributed from spurious background sources. Taking reported S/N ratios for $\lsat$ from T19, we compute mutual information between $x$ and $y$ as above and show a summary of mutual information values in Figure \ref{fig:MItest2}. Note that this figure shows mean values for 20 sets of random data, with standard errors.

%%%%%%%%%%%%%%%%%%%%%%%%%%%%%%%%%%%%%%%%%%%%%%%%%%

% Don't change these lines
\bsp	% typesetting comment
\label{lastpage}
\end{document}